\input harvmac
\input epsf
\input amssym
\noblackbox


\def\hat{\widehat}

\def\figin{\epsfcheck\figin}\def\figins{\epsfcheck\figins}
\def\epsfcheck{\ifx\epsfbox\UnDeFiNeD
\message{(NO epsf.tex, FIGURES WILL BE IGNORED)}
\gdef\figin##1{\vskip2in}\gdef\figins##1{\hskip.5in}
\else\message{(FIGURES WILL BE INCLUDED)}%
\gdef\figin##1{##1}\gdef\figins##1{##1}\fi}
\def\DefWarn#1{}
\def\figinsert{\goodbreak\midinsert}
\def\ifig#1#2#3{\DefWarn#1\xdef#1{fig.~\the\figno}
\writedef{#1\leftbracket fig.\noexpand~\the\figno}%
\figinsert\figin{\centerline{#3}}\medskip\centerline{\vbox{\baselineskip12pt
\advance\hsize by -1truein\noindent\footnotefont{\bf
Fig.~\the\figno:\ } \it#2}}
\bigskip\endinsert\global\advance\figno by1}


\lref\MeadeWD{
  P.~Meade, N.~Seiberg and D.~Shih,
  ``General Gauge Mediation,''
Prog.\ Theor.\ Phys.\ Suppl.\  {\bf 177}, 143 (2009).
[arXiv:0801.3278 [hep-ph]].
}

\lref\DvaliCU{
  G.~R.~Dvali, G.~F.~Giudice and A.~Pomarol,
  ``The Mu problem in theories with gauge mediated supersymmetry breaking,''
Nucl.\ Phys.\ B {\bf 478}, 31 (1996).
[hep-ph/9603238].
}

\lref\rattazi{
  G.~F.~Giudice, H.~D.~Kim and R.~Rattazzi,
  ``Natural mu and B mu in gauge mediation,''
  Phys.\ Lett.\  B {\bf 660}, 545 (2008)
  [arXiv:0711.4448 [hep-ph]].
}

\lref\BuicanWS{
  M.~Buican, P.~Meade, N.~Seiberg and D.~Shih,
  ``Exploring General Gauge Mediation,''
  JHEP {\bf 0903}, 016 (2009)
  [arXiv:0812.3668 [hep-ph]].
}

\lref\DumitrescuHA{
  T.~T.~Dumitrescu, Z.~Komargodski, N.~Seiberg and D.~Shih,
  ``General Messenger Gauge Mediation,''
  JHEP {\bf 1005}, 096 (2010)
  [arXiv:1003.2661 [hep-ph]].
}

\lref\KomargodskiAX{
  Z.~Komargodski and N.~Seiberg,
  ``mu and General Gauge Mediation,''
  JHEP {\bf 0903}, 072 (2009)
  [arXiv:0812.3900 [hep-ph]].
}

\lref\schmaltz{
  T.~S.~Roy and M.~Schmaltz,
  ``Hidden solution to the mu/Bmu problem in gauge mediation,''
Phys.\ Rev.\ D {\bf 77}, 095008 (2008).
[arXiv:0708.3593 [hep-ph]].
}

\lref\CraigXP{
  N.~Craig, S.~Knapen, D.~Shih and Y.~Zhao,
  ``A Complete Model of Low-Scale Gauge Mediation,''
[arXiv:1206.4086 [hep-ph]].
}

\lref\NillesDY{
  H.~P.~Nilles, M.~Srednicki and D.~Wyler,
  ``Weak Interaction Breakdown Induced by Supergravity,''
Phys.\ Lett.\ B {\bf 120}, 346 (1983)..
}

\lref\FrereAG{
  J.~M.~Frere, D.~R.~T.~Jones and S.~Raby,
  ``Fermion Masses and Induction of the Weak Scale by Supergravity,''
Nucl.\ Phys.\ B {\bf 222}, 11 (1983)..
}

\lref\LangackerHS{
  P.~Langacker, N.~Polonsky and J.~Wang,
  ``A Low-energy solution to the mu problem in gauge mediation,''
Phys.\ Rev.\ D {\bf 60}, 115005 (1999).
[hep-ph/9905252].
}

\lref\CsakiSR{
  C.~Csaki, A.~Falkowski, Y.~Nomura and T.~Volansky,
  ``New Approach to the mu-Bmu Problem of Gauge-Mediated Supersymmetry Breaking,''
Phys.\ Rev.\ Lett.\  {\bf 102}, 111801 (2009).
[arXiv:0809.4492 [hep-ph]].
}

\lref\YanagidaYF{
  T.~Yanagida,
  ``A Solution to the mu problem in gauge mediated supersymmetry breaking models,''
Phys.\ Lett.\ B {\bf 400}, 109 (1997).
[hep-ph/9701394].
}

\lref\DineSWA{
  M.~Dine and J.~Kehayias,
  ``Discrete R Symmetries and Low Energy Supersymmetry,''
Phys.\ Rev.\ D {\bf 82}, 055014 (2010).
[arXiv:0909.1615 [hep-ph]].
}

\lref\CraigRK{
  N.~J.~Craig and D.~Green,
  ``On the Phenomenology of Strongly Coupled Hidden Sectors,''
JHEP {\bf 0909}, 113 (2009).
[arXiv:0905.4088 [hep-ph]].
}

\lref\GreenNQ{
  D.~Green and D.~Shih,
  ``Bounds on SCFTs from Conformal Perturbation Theory,''
[arXiv:1203.5129 [hep-th]].
}

\lref\MurayamaGE{
  H.~Murayama, Y.~Nomura and D.~Poland,
  ``More visible effects of the hidden sector,''
Phys.\ Rev.\ D {\bf 77}, 015005 (2008).
[arXiv:0709.0775 [hep-ph]].
}

\lref\Weak{
N.~Craig, S.~Knapen and D.~Shih,
To appear.
}

\lref\futurework{
N.~Craig, S.~Knapen and D.~Shih,
To appear.
}

\lref\CraigXP{
  N.~Craig, S.~Knapen, D.~Shih and Y.~Zhao,
  ``A Complete Model of Low-Scale Gauge Mediation,''
[arXiv:1206.4086 [hep-ph]].
}

\lref\PolandEY{
  D.~Poland, D.~Simmons-Duffin and A.~Vichi,
  ``Carving Out the Space of 4D CFTs,''
JHEP {\bf 1205}, 110 (2012).
[arXiv:1109.5176 [hep-th]].
}

\lref\AllanachKG{
  B.~C.~Allanach,
  ``SOFTSUSY: a program for calculating supersymmetric spectra,''
Comput.\ Phys.\ Commun.\  {\bf 143}, 305 (2002).
[hep-ph/0104145].
}

\lref\BarbieriFN{
  R.~Barbieri and G.~F.~Giudice,
  ``Upper Bounds on Supersymmetric Particle Masses,''
Nucl.\ Phys.\ B {\bf 306}, 63 (1988)..
}

\lref\CMSmultilept{
 CMS Collaboration,
  ``Search for anomalous production of multilepton events in pp collisions at $\sqrt{s}$=7 TeV,''
Submitted to the Journal of High Energy Physics (2012).
[hep-ex/1204.5341].
}

\lref\DraperAA{
  P.~Draper, P.~Meade, M.~Reece and D.~Shih,
  ``Implications of a 125 GeV Higgs for the MSSM and Low-Scale SUSY Breaking,''
Phys.\ Rev.\ D {\bf 85}, 095007 (2012).
[arXiv:1112.3068 [hep-ph]].
}


\lref\ATLASHiggs{
  G.~Aad {\it et al.}  [The ATLAS Collaboration],
  ``Observation of a new particle in the search for the Standard Model Higgs boson with the ATLAS detector at the LHC,''
[arXiv:1207.7214 [hep-ex]].
}

\lref\CMSHiggs{
  S.~Chatrchyan {\it et al.}  [The CMS Collaboration],
  ``Observation of a new boson at a mass of 125 GeV with the CMS experiment at the LHC,''
[arXiv:1207.7235 [hep-ex]].
}

\lref\higgspapers{
  L.~J.~Hall, D.~Pinner and J.~T.~Ruderman,
JHEP {\bf 1204}, 131 (2012).
[arXiv:1112.2703 [hep-ph]].

  S.~Heinemeyer, O.~Stal and G.~Weiglein,
Phys.\ Lett.\ B {\bf 710}, 201 (2012).
[arXiv:1112.3026 [hep-ph]].

  A.~Arbey, M.~Battaglia, A.~Djouadi, F.~Mahmoudi and J.~Quevillon,
Phys.\ Lett.\ B {\bf 708}, 162 (2012).
[arXiv:1112.3028 [hep-ph]].

  A.~Arbey, M.~Battaglia and F.~Mahmoudi,
Eur.\ Phys.\ J.\ C {\bf 72}, 1906 (2012).
[arXiv:1112.3032 [hep-ph]].

  M.~Carena, S.~Gori, N.~R.~Shah and C.~E.~M.~Wagner,
JHEP {\bf 1203}, 014 (2012).
[arXiv:1112.3336 [hep-ph]].
}

\lref\Haber{
H.~E.~Haber and R.~Hempfling,
  ``Can the mass of the lightest Higgs boson of the minimal supersymmetric model be larger than m(Z)?,''
Phys.\ Rev.\ Lett.\  {\bf 66}, 1815 (1991)..

 R.~Barbieri and M.~Frigeni,
  ``The Supersymmetric Higgs searches at LEP after radiative corrections,''
Phys.\ Lett.\ B {\bf 258}, 395 (1991)..
}

\lref\stopmixing{
  J.~A.~Casas, J.~R.~Espinosa, M.~Quiros and A.~Riotto,
Nucl.\ Phys.\ B {\bf 436}, 3 (1995), [Erratum-ibid.\ B {\bf 439}, 466 (1995)].
[hep-ph/9407389].

  M.~S.~Carena, J.~R.~Espinosa, M.~Quiros and C.~E.~M.~Wagner,
Phys.\ Lett.\ B {\bf 355}, 209 (1995).
[hep-ph/9504316].

  H.~E.~Haber, R.~Hempfling and A.~H.~Hoang,
Z.\ Phys.\ C {\bf 75}, 539 (1997).
[hep-ph/9609331].

  F.~Brummer, S.~Kraml and S.~Kulkarni,
JHEP {\bf 1208}, 089 (2012).
[arXiv:1204.5977 [hep-ph]].
}

\lref\EllwangerAA{
  U.~Ellwanger,
JHEP {\bf 1203}, 044 (2012).
[arXiv:1112.3548 [hep-ph]].

  J.~F.~Gunion, Y.~Jiang and S.~Kraml,
Phys.\ Lett.\ B {\bf 710}, 454 (2012).
[arXiv:1201.0982 [hep-ph]].

  S.~F.~King, M.~Muhlleitner and R.~Nevzorov,
Nucl.\ Phys.\ B {\bf 860}, 207 (2012).
[arXiv:1201.2671 [hep-ph]].

}

\lref\BatraNJ{
  P.~Batra, A.~Delgado, D.~E.~Kaplan and T.~M.~P.~Tait,
  ``The Higgs mass bound in gauge extensions of the minimal supersymmetric standard model,''
JHEP {\bf 0402}, 043 (2004).
[hep-ph/0309149].

  A.~Arvanitaki and G.~Villadoro,
  ``A Non Standard Model Higgs at the LHC as a Sign of Naturalness,''
JHEP {\bf 1202}, 144 (2012).
[arXiv:1112.4835 [hep-ph]].

  M.~Endo, K.~Hamaguchi, S.~Iwamoto, K.~Nakayama and N.~Yokozaki,
  ``Higgs mass and muon anomalous magnetic moment in the U(1) extended MSSM,''
Phys.\ Rev.\ D {\bf 85}, 095006 (2012).
[arXiv:1112.6412 [hep-ph]].
}

\lref\DineDV{
  M.~Dine, P.~J.~Fox, E.~Gorbatov, Y.~Shadmi, Y.~Shirman and S.~D.~Thomas,
  ``Visible effects of the hidden sector,''
Phys.\ Rev.\ D {\bf 70}, 045023 (2004).
[hep-ph/0405159].
}

\lref\PerezNG{
  G.~Perez, T.~S.~Roy and M.~Schmaltz,
  ``Phenomenology of SUSY with scalar sequestering,''
Phys.\ Rev.\ D {\bf 79}, 095016 (2009).
[arXiv:0811.3206 [hep-ph]].
}

\lref\CohenQC{
  A.~G.~Cohen, T.~S.~Roy and M.~Schmaltz,
  ``Hidden sector renormalization of MSSM scalar masses,''
JHEP {\bf 0702}, 027 (2007).
[hep-ph/0612100].
}

\lref\KimSY{
  H.~D.~Kim and J.~-H.~Kim,
  ``Higgs Phenomenology of Scalar Sequestering,''
JHEP {\bf 0905}, 040 (2009).
[arXiv:0903.0025 [hep-ph]].
}

\lref\CraigVS{
  N.~J.~Craig and D.~R.~Green,
  ``Sequestering the Gravitino: Neutralino Dark Matter in Gauge Mediation,''
Phys.\ Rev.\ D {\bf 79}, 065030 (2009).
[arXiv:0808.1097 [hep-ph]].
}

\lref\AsanoQC{
  M.~Asano, J.~Hisano, T.~Okada and S.~Sugiyama,
  ``A Realistic Extension of Gauge-Mediated SUSY-Breaking Model with Superconformal Hidden Sector,''
Phys.\ Lett.\ B {\bf 673}, 146 (2009).
[arXiv:0810.4606 [hep-ph]].
}

\lref\DineYW{
  M.~Dine and A.~E.~Nelson,
Phys.\ Rev.\ D {\bf 48}, 1277 (1993).
[hep-ph/9303230].

  M.~Dine, A.~E.~Nelson and Y.~Shirman,
Phys.\ Rev.\ D {\bf 51}, 1362 (1995).
[hep-ph/9408384].

  M.~Dine, A.~E.~Nelson, Y.~Nir and Y.~Shirman,
Phys.\ Rev.\ D {\bf 53}, 2658 (1996).
[hep-ph/9507378].
}

\lref\HarnikRS{
  R.~Harnik, G.~D.~Kribs, D.~T.~Larson and H.~Murayama,
  ``The Minimal supersymmetric fat Higgs model,''
Phys.\ Rev.\ D {\bf 70}, 015002 (2004).
[hep-ph/0311349].
}

\lref\AzatovHT{
  A.~Azatov, J.~Galloway and M.~A.~Luty,
  ``Superconformal Technicolor,''
Phys.\ Rev.\ Lett.\  {\bf 108}, 041802 (2012).
[arXiv:1106.3346 [hep-ph]].
}

\lref\GiudiceNI{
  G.~F.~Giudice and R.~Rattazzi,
  ``Extracting supersymmetry breaking effects from wave function renormalization,''
Nucl.\ Phys.\ B {\bf 511}, 25 (1998).
[hep-ph/9706540].
}

\lref\KangRA{
  Z.~Kang, T.~Li, T.~Liu, C.~Tong and J.~M.~Yang,
 ``A Heavy SM-like Higgs and a Light Stop from Yukawa-Deflected Gauge Mediation,''
[arXiv:1203.2336 [hep-ph]].
}

\lref\CarpenterWI{
  L.~M.~Carpenter, M.~Dine, G.~Festuccia and J.~D.~Mason,
  ``Implementing General Gauge Mediation,''
Phys.\ Rev.\ D {\bf 79}, 035002 (2009).
[arXiv:0805.2944 [hep-ph]].
}

\lref\GiudiceNI{
  G.~F.~Giudice and R.~Rattazzi,
  ``Extracting supersymmetry breaking effects from wave function renormalization,''
Nucl.\ Phys.\ B {\bf 511}, 25 (1998).
[hep-ph/9706540].
}

\lref\DelgadoRZ{
  A.~Delgado, G.~F.~Giudice and P.~Slavich,
  ``Dynamical mu term in gauge mediation,''
Phys.\ Lett.\ B {\bf 653}, 424 (2007).
[arXiv:0706.3873 [hep-ph]].
}

\newfam\mbfcalfam
\font\tenmbfcal=cmbsy10
\font\sevenmbfcal=cmbsy7
\font\fivembfcal=cmbsy5
\textfont\mbfcalfam=\tenmbfcal
\scriptfont\mbfcalfam=\sevenmbfcal
\scriptscriptfont\mbfcalfam=\fivembfcal

\newfam\mscrfam
\font\tenmscr=rsfs10
\font\sevenmscr=rsfs7
\font\fivemscr=rsfs5
\textfont\mscrfam=\tenmscr
\scriptfont\mscrfam=\sevenmscr
\scriptscriptfont\mscrfam=\fivemscr


\def\CC{{\cal C}}

\def\CL{{\cal L}}

\def\CO{{\cal O}}

\def\CX{{\cal X}}


\def\m{\mu}

\def\l{\lambda}

\def\k{\kappa}

\def\D{\Delta}
\def\B{\hat B_\mu}

\def\dag{\dagger}

\def\Q{\bar Q}


\def\bra{\langle}
\def\ket{\rangle}

\def\lfm#1{\medskip\noindent\item{#1}}

\Title{RU-NHETC-2013-04}{General Messenger Higgs Mediation}


\centerline{Nathaniel Craig,$^{1,2}$ Simon Knapen,$^1$ and David Shih$^1$
}
\bigskip
\centerline{$^1${\it  Department of Physics and Astronomy, Rutgers University, Piscataway, NJ 08854, USA}}
\centerline{$^2${\it  School of Natural Sciences, Institute for Advanced Study, Princeton, NJ 08540, USA}}
\bigskip
\vskip 1cm

\noindent  We present a general formalism for analyzing supersymmetric models where the Higgs sector directly couples to the messengers of supersymmetry breaking. Such Higgs-messenger interactions are strongly motivated by the discovery of a Higgs boson near 125 GeV, but they also raise the specter of the $\mu/B_\mu$ and $A/m_H^2$ problems. Using our formalism, we identify new avenues to solving these problems through strong dynamics in the messenger sector or hidden sector. Although our formalism is entirely general, we show how it reproduces familiar results in two simplifying limits: one where the hidden sector consists of a single spurion, and the other where it is approximately superconformal. In the latter limit, our formalism generalizes and clarifies the scenario of hidden sector sequestering, which we show can solve both the $\mu/B_\mu$ and $A/m_H^2$ problems uniformly.

\bigskip

\Date{February 2013}

\newsec{Introduction}

The recent discovery \refs{ \ATLASHiggs, \CMSHiggs} of a Higgs-like particle with a mass near 125 GeV has profound implications for physics beyond the Standard Model. It renews the urgency of the hierarchy problem, for which supersymmetry (SUSY) remains the best solution. Minimal realizations of weak-scale SUSY such as the MSSM are highly constrained, since the tree-level prediction for the Higgs mass is bounded from above by the mass of the $Z$ boson and must be increased through radiative corrections. As discussed in \refs{\higgspapers, \DraperAA}, to obtain $m_h = 125$ GeV in the MSSM while minimizing the fine-tuning of the electroweak scale, the $A$-terms must be large relative to other soft masses and close to maximal mixing \refs{\stopmixing}. 

There are several options for generating large weak-scale $A$-terms in calculable models, and each comes with its own challenges. One option is to have $A\approx 0$ at the messenger scale $M$ but generate it through RG running of the MSSM. In \DraperAA,  it was shown that this places strong constraints on the messenger scale and the gluino mass, requiring both to be very high. Another option is to generate non-zero $A$-terms already at the messenger scale, by directly coupling $H_u$ and $H_d$ to the messengers.\foot{We could also consider direct couplings of the quark superfields to the messengers, but these would not be minimally flavor violating.} Aside from being richer in terms of model building possibilities, this option is attractive and economical because such couplings are already necessary for solving the $\mu$ problem of the MSSM. But here the main challenge, at least in weakly-coupled models, is something called the ``$A/m_H^2$ problem:" $A$ and $m_H^2$ are typically generated at the same loop order, in direct analogy with the $\mu/B_\mu$ problem \CraigXP. Such a large $m_H^2$ would have disastrous effects on electroweak symmetry breaking (EWSB) and fine-tuning. 

In \CraigXP, the problem of generating large $A$-terms was studied in the context of weakly-coupled messenger models  where SUSY is broken by a spurion $X$. Here the challenges of the $A/m_H^2$ problem are perhaps starkest. Integrating out the messengers generates effective operators involving the SUSY-breaking hidden sector and the Higgs fields; the $A$-terms arise from
\eqn\Aop{\eqalign{
 & c_{A_u}  \int d^4\theta\, {X^\dagger\over M} H_u^\dagger H_u \to A_u H^\dagger_u F_{H_u}
}}
where we have substituted $\langle X\rangle=\theta^2 F$. In general,  $m_{H_u}^2$ is also generated at the same loop order, since the only difference in the effective operator is the non-chiral operator $X^\dagger X$ instead of the chiral hidden sector operator $X$:
\eqn\mHop{\eqalign{
 & c_{m_{H_u}^2}\int d^4\theta\,  {X^\dagger X \over M^2} H_u^\dagger H_u \to  \hat m_{H_u}^2 H_u^\dagger H_u
}}
This is exactly analogous to the more well-known $\mu/B_\mu$ problem -- for which the effective operators are the same as in \Aop\ and \mHop, but with $H_u^\dagger H_u$ replaced by $H_u H_d$. In \CraigXP, it was argued (following \rattazi) that only models where the messengers receive all masses and SUSY-breaking from a single spurion (i.e.\ models of minimal gauge mediation (MGM) \DineYW) can solve the $A$/$m_{H}^2$ problem, by eliminating the one-loop $m_H^2$ at leading order in $F/M^2$. (See also \KangRA\ for a study of these MGM-based models.) But even in these models, a residual problem -- dubbed the ``little $A$/$m_{H}^2$ problem" in \CraigXP\ -- remains: $m_{H_u}^2$ always contains an irreducible, positive, two-loop contribution $\propto A_u^2$
coming from integrating out the auxiliary component of $H_u$ in \Aop. Since the $A$-terms must be large (at least $\sim 2$ TeV) for maximal stop mixing and $m_h=125$ GeV, this also presents difficulties for radiative EWSB and for fine tuning. 

Motivated by these considerations, in this paper we will broaden the scope of \CraigXP\ considerably and take a general, model-independent approach to studying the Higgs soft spectrum arising from direct Higgs-messenger couplings.  This will include both weakly-coupled spurion models and strongly-coupled models as special cases. Our main tool in this endeavor will be the supersymmetric correlator formalism of \refs{\MeadeWD,\BuicanWS}. This was first applied to the Higgs sector in \KomargodskiAX, assuming the following portals between the Higgs and hidden sectors \MeadeWD: 
\eqn\suppot{
W = \lambda_u\CO_u H_u+\lambda_d\CO_dH_d
}
where the hidden sector operators $\CO_{u,d}$ are $SU(2)$ doublets.\foot{Throughout the paper we will be neglecting potential contributions to soft masses proportional to the MSSM gauge and Yukawa couplings. Note that the presence of hidden sector operators $\CO_{u, d}$ implies at the minimum some gauge-mediated contributions to the MSSM soft spectrum, but these need not be the leading effect; the interactions in \suppot\ may be incorporated into various models of supersymmetry breaking. } (Singlet couplings were also studied in \KomargodskiAX; in the interest of clarity we will only work out the doublet case in this paper. The extension to the singlet case is straightforward.)  We will refer to this framework as ``General Higgs Mediation" (GHM), in analogy with \refs{\MeadeWD,\BuicanWS}.  Integrating out the hidden sector generates  the Higgs-sector soft Lagrangian:\foot{Here and for the rest of the paper, we are neglecting the ``wrong Higgs couplings," as they arise at a higher order in the supersymmetry breaking order parameter \KomargodskiAX. }
\eqn\conpriorFint{\eqalign{- \delta\CL &\supset \Big(A_{u} H^\dagger_{u} F_{H_{u}}+A_{d} H^\dagger_{d} F_{H_{d}}+c.c.\Big)-\Big({\hat m}_{H_{u}}^2 H_{u}^\dagger H_{u}+{\hat m}_{H_{d}}^2 H_{d}^\dagger H_{d}\Big)
\cr
&
\quad +\m\Big(H_u F_{H_d}+H_dF_{H_u}-\psi_{H_u} \psi_{H_d}+c.c. \Big)
 -\Big( \B H_uH_d + c.c.\Big)
}} 
Note that we have put hats on the dimension-two soft masses in \conpriorFint, in order to distinguish them from the full dimension-two soft masses that are only obtained upon integrating out $F_{H_{u,d}}$:
\eqn\dimtwofull{
m_{H_{u,d}}^2 = \hat m_{H_{u,d}}^2 + |A_{u,d}|^2,\qquad B_\mu = \hat B_\mu + \mu(A_u^*+A_d^*)
}
Correlator formulas for the Higgs soft parameters were derived to leading order in $\lambda_{u,d}$ in \KomargodskiAX. No assumption was made in \KomargodskiAX\ regarding the structure of the hidden sector, thus their results were best suited to the single-sector case where there is no distinction between messenger and SUSY-breaking hidden sectors. 

Here we will extend the work of \KomargodskiAX\ in two ways. First, we will extend their single-sector formulas for the dimension-two soft masses to next-to-leading order in $\lambda_{u,d}$. This is obviously necessary in order to discuss phenomenologically relevant models where $A_{u,d}^2\sim m_{H_{u,d}}^2$ and $B_\mu \sim\mu^2$. Second, we will derive more detailed formulas for models in which the messenger sector is distinct from the SUSY-breaking hidden sector, along the lines of \DumitrescuHA. This factorization is illustrated in fig.\ 1, and in analogy with \DumitrescuHA, we will refer to this framework as ``General Messenger Higgs Mediation" (GMHM). We will focus on the superpotential portal of \DumitrescuHA:
\eqn\suppotportal{
W = \k\CO_h\CO_m
}
In general, the coupling $\kappa$ can be dimensionful, and $\CO_{h,m}$ have dimensions $\Delta_{h,m}$. $\CO_h$ is a chiral operator that breaks SUSY
\eqn\Fdef{
\langle Q^2 \CO_h\rangle \equiv F^{\Delta_h+1\over 2}
}
and generalizes the spurion $X$ to possibly nontrivial, interacting hidden sectors. (Without loss of generality, we take $F$ to be real, and we shift $\CO_h$ so that its lowest component has zero vacuum expectation value.) It would also be interesting to study the K\"{a}hler and half-K\"{a}hler portals considered in \DumitrescuHA, but we will not do so here. 

One of the primary virtues of GMHM is that it enables the study of models where the SUSY-breaking scale $\sqrt{F}$ is much smaller than the messenger scale $M$. When this is the case, there is an additional small parameter $F/M^2$ to expand in, and the expressions for the soft masses often simplify. More generally, many existing models feature this separation between SUSY-breaking hidden sector and messenger sector, and so GMHM is the ideal framework for studying them collectively.

\ifig\gmgmsetup{The general setup of GMHM, assuming doublet portals connecting the Higgs sector to the messenger sector. The messengers are characterized by a scale $M$, and they communicate via another perturbative superpotential interaction with the hidden sector, which is characterized by a SUSY-breaking scale $\sqrt{F}$.}{\epsfxsize=0.85\hsize\epsfbox{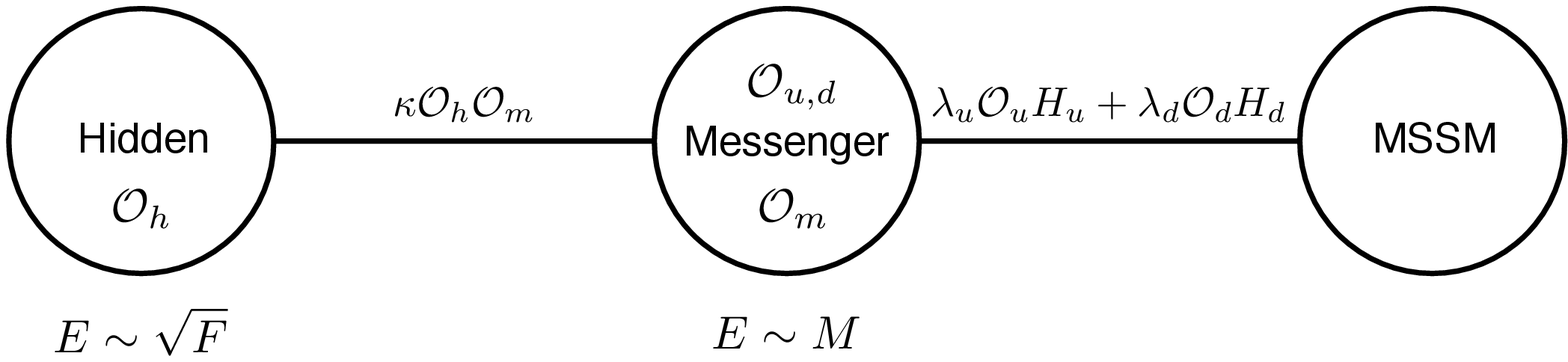}}

Computing soft parameters in the framework of GMHM involves a double expansion in $\lambda_{u,d}$ and $\kappa$. Carefully performing this double expansion and manipulating the resulting correlators, we will derive fully general formulas for Higgs soft parameters in any setup of the form in fig.\ 1:
\eqn\softparamgmgmintro{\eqalign{
\mu &=\lambda_u \lambda_d  \kappa^* \, \langle \bar Q^2 \CO_h^\dag \rangle_h \int d^4y\, C_\mu(y)\cr
A_{u,d} &= |\lambda_{u,d}|^2 \kappa^* \,\langle \bar Q^2 \CO_h^\dag \rangle_h \int d^4y\, C_{A_{u,d}}(y)\cr
B_{\mu} &=\lambda_u\lambda_d |\kappa|^2\, \int d^4yd^4y'\,\langle Q^4[\CO_h^\dagger(y) \CO_h(y')]\rangle_h C_{B_\mu}(y,y';\lambda_{u,d})\cr
m_{H_{u,d}}^2 &=-|\mu|^2 + |\lambda_{u,d}|^2 |\kappa|^2\, \int d^4yd^4y' \,\langle Q^4[\CO_h^\dagger(y) \CO_h(y')]\rangle_h C_{m_{H_{u,d}}^2}(y,y';\lambda_{u,d})
}}
where $Q^4= Q^2\bar Q^2$. $C_\mu$, etc.\ are integrated correlation functions of messenger-sector operators; explicit expressions for them will be given in Section 2. Since we have expanded to NLO in $\lambda_{u,d}$, $C_{B_\mu}$ and $C_{m_{H_{u,d}}^2}$ contain $\CO(|\lambda_{u,d}|^2)$ corrections.

 These formulas have broad applicability, as they may be used to compute Higgs soft parameters for any model with Higgs-messenger couplings in which the messenger sector and SUSY-breaking hidden sector factorize. We will illustrate this in  several ways, starting with showing how they reproduce the results of the weakly-coupled spurion models of \CraigXP. In these models, the hidden sector has no dynamics, and so 
 \eqn\spurionintro{
 \langle Q^4[\CO_h^\dagger(y) \CO_h(y')]\rangle_h\to |\langle Q^2 \CO_h\rangle|^2
 } 
We will show how the $A/m_H^2$ problem is a generic property of the integrated messenger correlators $\int C_{m_{H_{u,d}}^2}$ and $\int C_{A_{u,d}}$, and how the little $A/m_H^2$ problem (made explicit in \dimtwofull) arises from the disconnected part of $C_{m_{H_{u,d}}^2}$.
  
The utility of the GMHM framework extends far beyond spurion-messenger models, however. 
Such broadening of scope is powerfully motivated by the challenges that weakly-coupled spurion models face in accommodating the Higgs mass. Perhaps the key lies in non-trivial dynamics of the hidden sector, as the considerations that led to the $A/m_H^2$ problem could be completely avoided at strong coupling. For instance, if the messenger sector is strongly-coupled, the notion of a loop factor might not even apply. Or, as has been suggested before in the context of the $\mu/B_\mu$ problem  \refs{\DineDV, \MurayamaGE, \schmaltz}, if the SUSY-breaking hidden sector is strongly coupled, then $\CO_h^\dagger \CO_h$ should really be replaced with a general non-chiral operator $\CO_\Delta$ with scaling dimension $\Delta$. If $\Delta>2\Delta_h$ and $\sqrt{F}\ll M$, then the anomalous dimension of $\CO_\Delta$ could help to sequester $B_\mu$ relative to $\mu^2$: 
\eqn\muBmuseq{
{B_\mu\over \mu^2} \sim \left({\sqrt{F}\over M}\right)^{\Delta-2\Delta_h}\ll 1
}
 It is natural to ask whether the same mechanism can help with the $A/m_H^2$ problem. As we will see, GMHM is ideally suited to addressing such questions. We will show that hidden-sector sequestering is contained within the GMHM framework, and that it can be successfully applied to both the $\mu/B_\mu$ and the $A/m_H^2$ problems. In particular, we will demonstrate how dependence on the hidden sector OPE and anomalous dimensions emerges naturally from \softparamgmgmintro.

In the course of generalizing hidden-sector sequestering using GMHM, we will clear up a lingering disagreement regarding the sequestered soft spectrum. In \refs{\MurayamaGE,\PerezNG}, it was claimed that the result of complete hidden-sector sequestering should be:
\eqn\hidsecseqclaimi{
m_{H_{u,d}}^2 \to  -\mu^2\,\, ,\quad B_\mu \to 0
}
In particular, the fully sequestered soft parameters do not depend on the $A$-terms, nor do they depend on OPE coefficients in the hidden sector. The claim was based on an argument that the $A$-term and $\mu$-term operators were redundant, in the sense that they could be removed by a field redefinition \PerezNG. 

Various questions were raised in \CraigRK\ about the validity of this argument -- what if the UV theory is strongly coupled and field redefinitions are not well-defined? If the UV theory is an interacting SCFT, shouldn't the OPE coefficient of $\CO_h^\dagger \CO_h\to\CO_\Delta$ be involved?  Using superconformal perturbation theory, \CraigRK\ argued that the result of hidden sector sequestering, starting from an interacting SCFT in the UV, should really be:
\eqn\hidsecseqclaimii{
m_{H_{u,d}}^2 \to  -\CC_\Delta\mu^2+(1-\CC_\Delta)A_{u,d}^2\,\, ,\quad B_\mu \to (1-\CC_\Delta)\mu(A^\ast_u+A^\ast_d)
}
where $\CC_\Delta$ is an OPE coefficient. This obviously differs from \hidsecseqclaimi. 

These previous studies of nontrivial hidden sector dynamics have all been based on RG evolution in the effective theory below the messenger scale. As we will see, in GMHM we instead work with the full theory and expand systematically in the couplings, expressing everything in terms of integrals over correlation functions. (In this sense GMHM is like a fixed-order calculation vs.\ the ``running and matching" taken in previous works.) This allows for more precise control over the final answer and a clearer understanding of the interplay between different contributions.
Using our general GMHM formulas, valid for any SUSY-breaking hidden sector and any messenger sector, we will show that -- surprisingly -- GMHM reproduces the claims of \refs{\MurayamaGE, \PerezNG}  and \hidsecseqclaimi, {\it even in strongly coupled cases where field redefinitions are not necessarily applicable}. We will reconcile the conformal perturbation theory RGEs derived in \CraigRK\ with \hidsecseqclaimi, vis a vis an approximate sum rule derived from the OPE. 

Significantly, applying the GMHM formalism to models of hidden sector sequestering allows us to go beyond simply clarifying existing results. In particular, the case of complete sequestering advocated in \refs{\MurayamaGE, \PerezNG} is an idealized limit in which $M \gg \sqrt{F}$ and $\Delta \gg 2 \Delta_h$. However, phenomenological considerations \refs{\PerezNG, \AsanoQC}\  and recent bounds on operator dimensions \PolandEY\ constrain these respective inequalities, so that viable models are only partially sequestered and remain sensitive to the details of the hidden sector. As we will show, the GMHM formalism provides an efficient framework for computing the soft spectrum of such partially sequestered models.

 The outline of our paper is as follows: In Section 2 we apply the GMHM formalism to the Higgs sector and obtain general NLO expressions for Higgs soft parameters given the portals  \suppot\ and \suppotportal.  We demonstrate their power in Section 3 by computing Higgs soft parameters in the spurion limit and the SCFT limit.  
In Section 4 we connect our GMHM results to previous work on hidden sector sequestering by computing Higgs soft parameters in an effective theory framework. We find perfect agreement between our GMHM results and various methods for computing soft parameters in the effective theory. In the process we reconcile results from superconformal perturbation theory with GMHM through an approximate sum rule derived from the OPE. We reserve various technical details of the GMHM framework for Appendix A. In Appendix B, we describe a check of the superconformal perturbation theory RGEs and the validity of field redefinitions using a perturbative Banks-Zaks fixed point.

\newsec{The Higgs in GMHM}

\subsec{General Higgs Mediation at NLO}

In this section, we will derive correlator formulas for the Higgs soft parameters in GMHM.  The first step is to expand in the direct Higgs-hidden-sector couplings \suppot, assuming a fully general hidden sector. In \KomargodskiAX, this was performed to leading order in $\lambda_{u,d}$. Since a successful solution to the $\mu/B_\mu$ and $A/m_H^2$ problems will have $B_\mu\lesssim\mu^2$ and $m_{H_u}^2\lesssim A_u^2$, we must extend the results of \KomargodskiAX\ by going to NLO for the dimension-two soft masses. We find:
\eqn\softparamdoublet{\eqalign{
  \mu &= \lambda_u\lambda_d\langle \CX_\mu\rangle\cr
  A_{u,d} &= |\lambda_{u,d}|^2 \langle \CX_{A_{u,d}}\rangle\cr
  \hat B_\mu &=\lambda_u\lambda_d \langle \CX_{B_\mu}\rangle\cr
  \hat m_{H_{u,d}}^2 &= |\lambda_{u,d}|^2\langle \CX_{m_{H_{u,d}}^2}\rangle
 }}
where we have introduced the following notation for later convenience:\foot{A note about our slightly non-standard conventions for the supercharges $Q_\alpha$ and $\bar Q_{\dot\alpha}$. To avoid cluttering our formulas with irrelevant factors of two, we are normalizing $Q$ and $\bar Q$ so that for a WZ model, $-\CL= Q^4 K + (Q^2 W + c.c.)$. This differs from the more standard conventions of e.g.\ Wess and Bagger that would have $1/16$ and $1/4$ in front of the K\"ahler potential and superpotential respectively.}   
\eqn\Xdef{\eqalign{
 \CX_\mu &=  -\int d^4x\; Q^\alpha \CO_u(x)Q_\alpha \CO_d(0) \cr
  \CX_{A_{u,d}} &= +\int d^4x\;\Q^2 \Big[\CO_{u,d}(x) \CO^\dagger_{u,d}(0)\Big] \cr
 \CX_{ B_\mu  } &=  -\int d^4x\;Q^2 \CO_u(x)Q^2 \CO_d(0) \left( 1 +  \sum_{i=u,d} |\lambda_i|^2\int d^4z\,d^4z'Q^2[ \CO_iH_i(z)]\bar Q^2[\CO_i^\dagger H_i^\dagger(z')]\right) \cr
 \CX_{ m_{H_{u,d}}^2 } &= -\int d^4x\;Q^2\Q^2 \Big[\CO_{u,d}(x) \CO^\dagger_{u,d}(0)\Big] \left(1 +  \sum_{i=u,d} |\lambda_i|^2\int d^4z\,d^4z'Q^2[ \CO_iH_i(z)]\bar Q^2[\CO_i^\dagger H_i^\dagger(z')]\right)\cr
}}
Note that we Wick rotated the formulas from \KomargodskiAX\ to Euclidean space, to avoid a proliferation of factors of $i$.

\ifig\GGMdisc{Topologies for the NLO expansion of the dimension-two soft parameters. The dashed lines, the solid line and the double lines represent the Higgs scalar, fermion and auxiliary propagators, respectively. The shaded blobs represent connected hidden-sector correlators. The diagrams on the bottom line are not 1PI and therefore do not contribute to $\hat B_\mu$ and $\hat m^2_{H_{u,d}}$. Note that there is no diagram of this type with an intermediate fermion line, as the individual correlators would have to be Grassmann odd.}{\epsfxsize=0.85\hsize\epsfbox{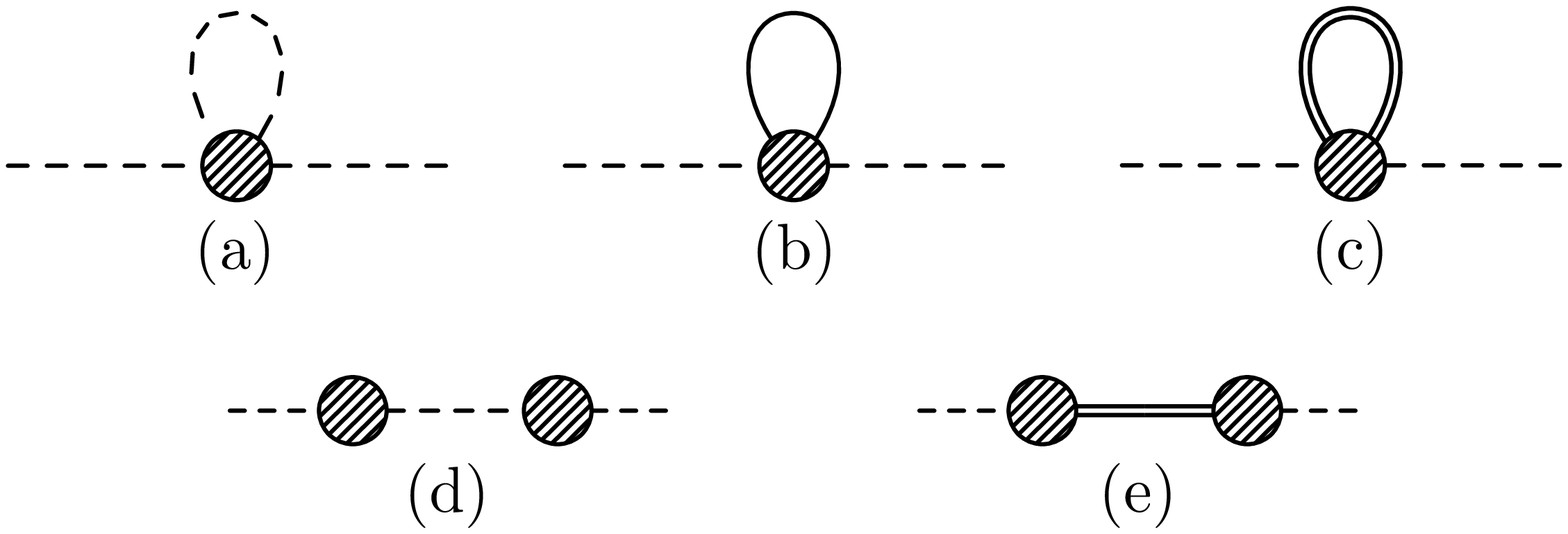}}

Since we are computing terms in an effective action, all diagrams contributing to \softparamdoublet\ must be 1PI. This becomes an issue  first at NLO order in $\l_{u,d}$, where we must contract the extra Higgs fields in the last two lines of \Xdef. Shown in fig.~2 are the different topologies for the diagrams at this order. Each blob is a connected (or if necessary, 1PI) hidden-sector correlator. The bottom two diagrams are interesting, since they involve {\it disconnected} hidden sector correlators. Let's now discuss  these topologies in turn:

\item{1.} Clearly, the three topologies\foot{An interesting subtlety about diagram (c): since the LO contribution is non-vanishing, the NLO contribution might be scheme dependent. In particular, employing the $\delta$-function from contracting the Higgs $F$-components in \Xdef\ collides the operators $\CO_i$ and $\CO_i^\dagger$, which can generate a UV divergence.} (a), (b), and (c) should always be included in the calculation of $m_{H_{u,d}}^2$ and $B_\mu$. 

\item{2.}  The topology (d) should not contribute, since it is not 1PI. Note that these diagrams are always schematically of the form $m_1^2 \times {1\over z^2} \times m_2^2$ where $m_{1,2}^2$ are shorthand for $B_\mu$, $m_{H_{u,d}}^2$. So if the soft masses are further suppressed by an additional small parameter (such as the GMHM portal $\kappa$), then this topology will always be higher order in this parameter. 

\item{3.} Finally, the topology (e) is not 1PI in the theory \conpriorFint\ that includes the Higgs auxiliary fields. However, these auxiliary fields must be integrated out, and the full dimension-two soft masses are given by \dimtwofull. This corresponds precisely to adding back in the topology (e) of \GGMdisc. For instance, taking the NLO contribution to $\langle\CX_{B_\mu}\rangle$ with $i=u$, contracting the auxiliary Higgs propagators, and disconnecting the correlator, we obtain:
\eqn\disconngen{\eqalign{ 
\int \delta^{(4)}(z-z') \Big \langle   Q^2 \CO_u(x)Q^2 \CO_d(0) \, \CO_u(z) \CO_u^\dagger(z')\Big\rangle  \to   \int \Big\langle Q^2 \CO_u(x) \CO_u^\dagger(z) \Big\rangle \Big\langle Q^2\CO_d(0)\CO_u(z)\Big\rangle \cr
}}
The integral on the right can be fully factorized using the translation invariance of both correlators and a simple change of variables. After putting back in all the couplings etc., this becomes $ A_u^* \mu$. A complete set of such disconnected diagrams is shown in fig.~3. Taking all of these into account exactly reproduces \dimtwofull.
\medskip

To summarize, when computing the full $B_\mu$ and $m^2_{H_{u,d}}$, we should in fact include diagrams of the type (e) in \GGMdisc, despite the fact that they do not appear to be 1PI at first glance. Meanwhile, disconnected correlators connected by a scalar propagator as in topology (d) must still be excluded from the NLO formulas.

\vskip0.5in

\ifig\GGMwithCuts{The NLO contributions to $B_\mu$ (upper two) and $m^2_{H_{u}}$ (lower two) involving contractions of the auxiliary fields of the Higgs multiplets. The contributions to $m^2_{H_d}$ are identical to $m^2_{H_{u}}$ upon switching $u\leftrightarrow d$ everywhere. When cut at the dotted line, these diagrams provide the extra contributions in \dimtwofull.}{\epsfxsize=0.85\hsize\epsfbox{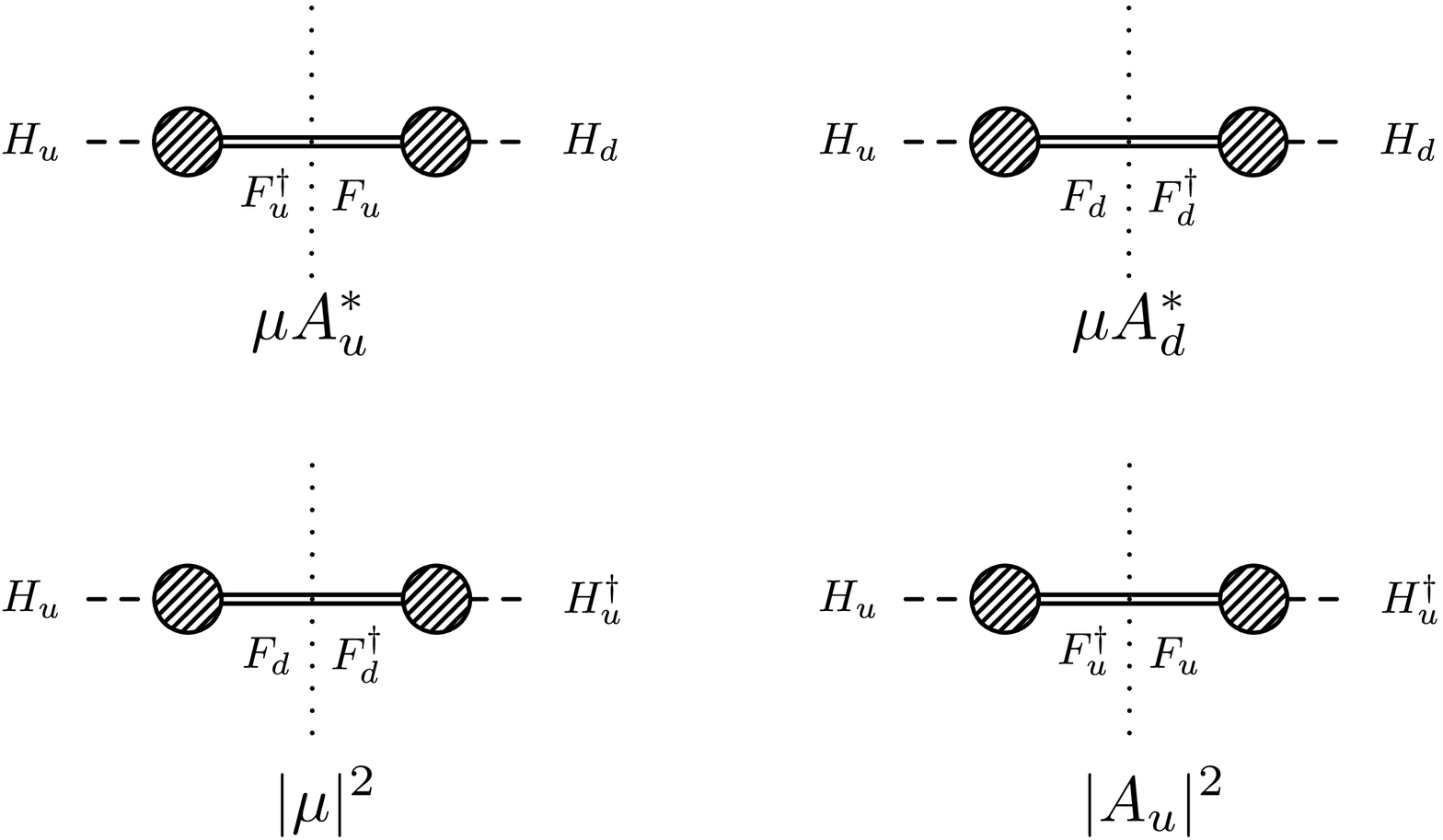}}

\subsec{Higgs soft parameters in GMHM} 

As discussed in the introduction, in the GMHM setup we further divide the overall hidden sector into a separate SUSY-breaking hidden sector and messenger sector, connected by a weakly-coupled portal. (We take the operators $\CO_u$ and $\CO_d$ to be in the messenger sector, as shown in \gmgmsetup.) Although in \DumitrescuHA\ more general portals were considered, in this paper we are focusing on the superpotential portal \suppotportal\ for simplicity. We then expand in $\kappa$ and factorize the correlators \softparamdoublet\ into separate correlators of the messenger and hidden sectors. Supersymmetry in the messenger sector then allows us to simplify the resulting expressions.

One general problem that immediately arises is that one typically finds both dimension-one soft masses and $B_\mu$ already at $\CO(\kappa)$. This would be disastrous for EWSB, as it would imply $B_\mu\sim \mu \times M$, where $M$ is the messenger scale. As we discuss more in Appendix A, a symmetry of the messenger sector that can forbid this while allowing for nonzero $\mu$ and $A_{u,d}$ (and gaugino masses) is an $R$-symmetry under which
\eqn\Rsymmetry{
 R(\CO_m)=2  , \quad R(\CO_u)+R(\CO_d)=4
 }
We will assume this $R$-symmetry throughout the paper.

With this in hand, we find that the GHM expressions \softparamdoublet\ become, at the leading nonvanishing order in $\kappa$: 
\eqn\softparamgmgm{\eqalign{
\mu &=\lambda_u \lambda_d  \kappa^* \,  \langle \bar Q^2 \CO_h^\dag \rangle_h \int d^4y\, \Big\langle\CO_m^\dagger(y)\CX_\mu\Big\rangle_m\cr
A_{u,d} &= |\lambda_{u,d}|^2 \kappa^* \, \langle \bar Q^2 \CO_h^\dag \rangle_h \int d^4y\, \Big\langle\CO_m^\dagger(y)\CX_{A_{u,d}}\Big\rangle_m\cr
\hat B_{\mu} &=\lambda_u\lambda_d |\kappa|^2\, \int d^4y\,d^4y'\,\Big\langle Q^4\Big[ \CO_h^\dagger(y)\CO_h(y')\Big]\Big\rangle_h \,\Big\langle\CO_m(y)\CO_m^\dagger(y')\CX_{B_\mu}\Big\rangle_m\cr
\hat m_{H_{u,d}}^2 &=|\lambda_{u,d}|^2 |\kappa|^2\, \int d^4y\,d^4y' \Big\langle Q^4\Big[ \CO_h^\dagger(y)\CO_h(y')\Big]\Big\rangle_h\, \Big\langle \CO_m(y)\CO_m^\dagger(y')\CX_{m_{H_{u,d}}^2}\Big\rangle_m\cr
}}
For more details, we refer the reader to Appendix A. Here the $m$ and $h$ subscripts denote correlators evaluated purely in the messenger and SUSY-breaking hidden sector, respectively. The integrated operators $\CX_\mu$ etc.\ were defined in \Xdef; now the components of the Higgs fields are understood to be contracted. In the last two lines we see that the answers always organize themselves so that they depend on a single hidden sector correlator, $\left\langle Q^4\Big[ \CO_h^\dagger(y)\CO_h(y')\Big]\right\rangle_h$. 

\ifig\GMHMtopologies{Possible topologies for the NLO in $\lambda_{u,d}$ contributions to the dimension-two soft parameters in GMHM. Blobs denoted with m (h) denote messsenger (hidden sector) correlators. The thick solid lines in (1) and (2) represent the sum of scalar, fermionic and auxiliary Higgs propagators.}{\epsfxsize=0.7\hsize\epsfbox{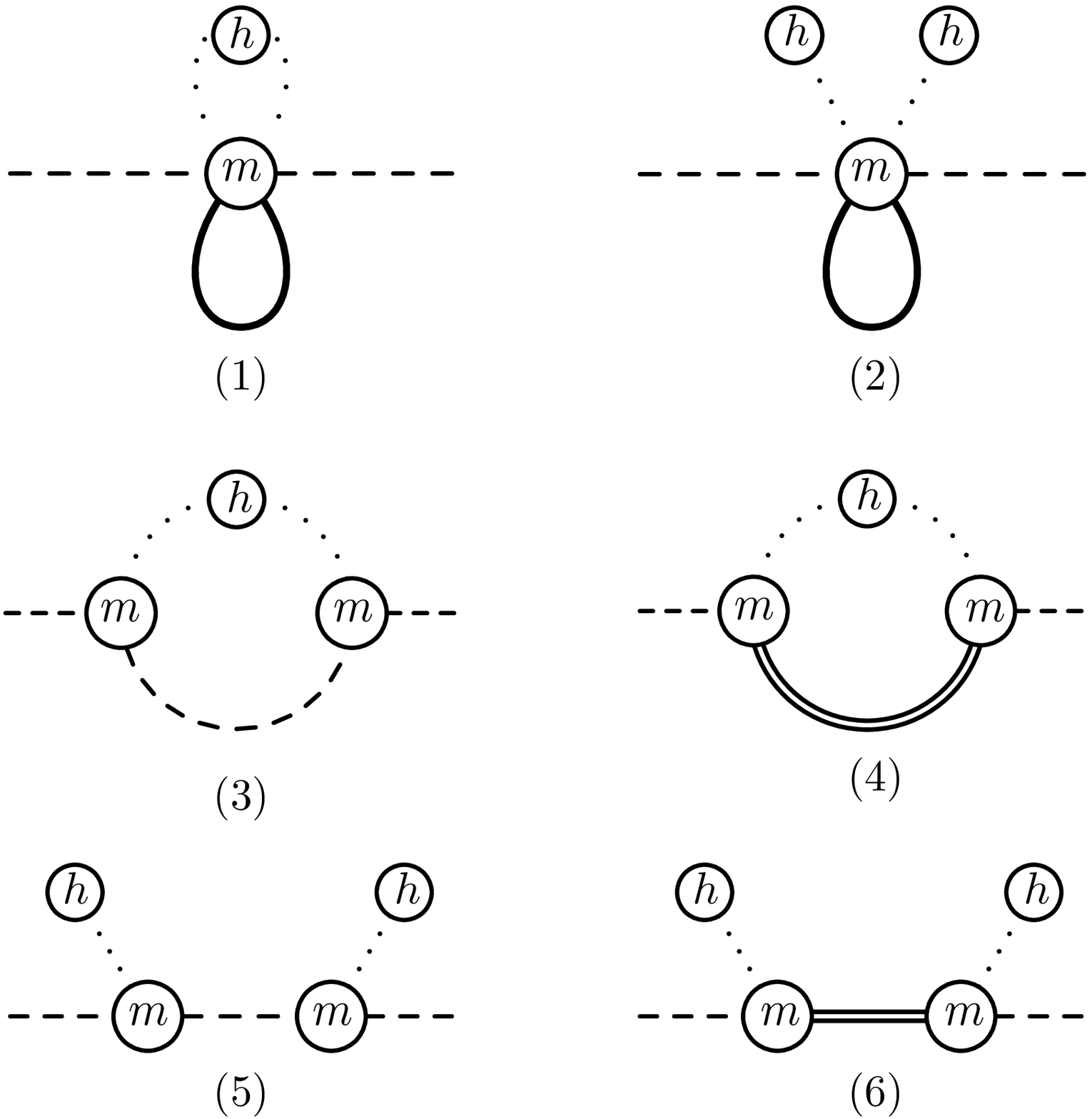}}

As in the previous subsection, at NLO in $\lambda_{u,d}$, we must again deal with the issue of connected vs.\ disconnected correlators. The NLO topologies in GMHM are shown in \GMHMtopologies, in direct analogy with \GGMdisc. As argued in the previous subsection, topology (6) is the contribution \dimtwofull\ of integrating out $F_{H_{u,d}}$, so it must be included in the final result for $B_\mu$ and $m_{H_{u,d}}^2$. Due to the $R$-symmetry and supersymmetry, topologies (3) and (5) do not contribute. Finally, the other topologies must clearly be included since they are 1PI. Therefore, we conclude that {\it the full $B_\mu$ and $m_{H_{u,d}}^2$ are given by the full hidden and messenger correlators, to this order in the GMHM expansion.} The final formulas are thus simply
\eqn\softparamgmgmdimtwofinal{\eqalign{
 B_{\mu} &=\lambda_u\lambda_d |\kappa|^2\, \int d^4y\,d^4y'\,\Big\langle Q^4\Big[ \CO_h^\dagger(y)\CO_h(y')\Big]\Big\rangle_{h,full} \,\Big\langle\CO_m(y)\CO_m^\dagger(y')\CX_{B_\mu}\Big\rangle_{m,full}\cr
 m_{H_{u,d}}^2 &=-|\mu|^2 + |\lambda_{u,d}|^2 |\kappa|^2\, \int d^4y\,d^4y' \Big\langle Q^4\Big[ \CO_h^\dagger(y)\CO_h(y')\Big]\Big\rangle_{h,full}\, \Big\langle \CO_m(y)\CO_m^\dagger(y')\CX_{m_{H_{u,d}}^2}\Big\rangle_{m,full}\cr
}}
where for $m_{H_{u,d}}^2$ we have subtracted out $|\mu|^2$ to adhere to the standard convention for these soft masses. These formulas are valid at $\CO(|\kappa|^2)$ and up to $\CO(|\lambda_{u,d}|^4)$, i.e.\ at the same order in the GMHM expansion as our results for $\mu^2$, etc. 

Let us conclude this section with one important observation about  \softparamgmgmdimtwofinal\ that we will need later: even though the full messenger correlators -- including disconnected parts -- are used in \softparamgmgmdimtwofinal, in fact only the region of integration with $|y-y'|\lesssim 1/M$ contributes to the soft masses. The reason is that the full messenger correlators fall off exponentially at long distance: 
\eqn\falloff{
\Big\langle \CO_m(y)\CO_m^\dagger(y')\CX_{B_\mu, m_{H_{u,d}}^2}\Big\rangle_{m,full} \to 0 \; \; {\rm as} \; \; |y-y'| \gg 1/M
}
since effectively only connected messenger diagrams contribute  after integrating out the Higgs auxiliary fields.  (For more explicit details, we again refer the reader to Appendix A.) This implies that when $\sqrt{F}\ll M$ -- as is generally the case in models of dynamical SUSY breaking -- the hidden sector correlator is effectively at short distance and the expressions \softparamgmgmdimtwofinal\ can be further simplified using the OPE in the hidden sector. We will put this observation to work in the next section when we discuss hidden sectors that are approximately superconformal  at the scale $M$.

\newsec{Examples}

The power of the GMHM formalism becomes apparent upon considering various special cases in which the general expressions \softparamgmgm\  simplify further. As was shown in \DumitrescuHA, illustrative examples include the well-known spurion limit employed in the study of many weakly-coupled models (such as those in \refs{\CraigXP,\KangRA}); and the SCFT limit used to study hidden sector sequestering  \refs{\DineDV\MurayamaGE\schmaltz\PerezNG \CraigRK-\AsanoQC}. As we will see, the  latter idea is especially attractive -- although originally proposed for solving the $\mu/B_\mu$ problem, we will show that it can work equally well for the $A/m_H^2$ problem. In the following subsections, we will consider these two special limits in turn, and show how they are reproduced in the GMHM framework.

\subsec{Spurion limit}

In the spurion limit, the hidden sector operator $\CO_h$ has no nontrivial interactions, and all hidden-sector correlators are given by their fully disconnected components.  Although it is not necessary for the spurion limit, for simplicity, we will take $\CO_h$ to have canonical dimension $\Delta_h=1$ in this subsection. So we have:
\eqn\Fvev{
\langle Q^2 \CO_h\rangle_h = F\, , \qquad \langle Q^4 \big[\CO_h(x)\CO_h^\dagger(0)\big]\rangle_{h,full} =|\langle Q^2 \CO_h\rangle_h|^2= |F|^2
}

The formulas for $\mu$ and $A_{u,d}$ are identical to those in \softparamgmgm. For $B_\mu$ and $m_{H_{u,d}}^2$, we saw in the previous section that the fully disconnected contributions (i.e.\ disconnecting both hidden and messenger correlators) are precisely those of integrating out $F_{H_{u,d}}$ as in \dimtwofull. Thus from \softparamgmgmdimtwofinal, we have: 
\eqn\softfinalspurion{\eqalign{
B_\mu &= \mu(A^\ast_u+A^\ast_d) + \lambda_u\lambda_d |\kappa|^2  |F|^2
\int d^4y\,d^4y' \, \Big\langle \CO_m(y)\CO_m^\dagger(y')\CX_{B_\mu}\Big\rangle_{m,connected}\cr
m_{H_{u,d}}^2  &= |A_{u,d}|^2 +   |\lambda_{u,d}|^2 |\kappa|^2  |F|^2 \int d^4y\,d^4y' \, \Big\langle \CO_m(y)\CO_m^\dagger(y')\CX_{m_{H_{u,d}}^2}\Big\rangle_{m,connected}
}}
Here we see that whether there is a $\mu/B_\mu$ or $A/m_{H}^2$ problem depends on the messenger sector in the following manner: 

\lfm{1.} If the messenger sector is strongly coupled, then loop counting is not well-defined, and there need not be any problem with $\mu/B_\mu$ or $A/m_{H}^2$.  However, there is not much more that we can say about this scenario, since the messenger sector is strongly coupled and typically incalculable, and statements about the parametric form of the soft masses are exhausted by dimensional analysis.

\lfm{2.} If instead the messenger sector is weakly coupled, then the messenger correlators can be computed, and they generally include a loop factor ($1/16\pi^2$) in addition to dimensional analysis. For generic messenger sectors, the connected correlators in \softfinalspurion\ are non-zero at one loop, which results in
\eqn\softparamweak{
\m \sim {\lambda_u \lambda_d \over 16 \pi^2} {F \over M}  \qquad B_\mu \sim {\lambda_u \lambda_d \over 16 \pi^2} {F^2 \over M^2}  \qquad A_{u,d} \sim {|\lambda_{u,d}|^2 \over 16 \pi^2} {F \over M}  \qquad m_{H_{u,d}}^2 \sim {|\lambda_{u,d}|^2 \over 16 \pi^2} {F^2 \over M^2}  .}
Now the $\mu/B_\mu$ and $A/m_{H}^2$ problems are manifest. So, in contrast to the strongly-coupled case, a weakly coupled messenger sector typically implies the existence of a  $\mu/B_\m$ and $A/m_{H}^2$ problem. 

\medskip

The $A/m_H^2$ problem is especially manifest in \softfinalspurion, since using \Xdef\ we can rewrite the LO formulas for $A$ and $m_H^2$ as:
\eqn \AmhProblemdoublet{\eqalign{
&A_{u,d}=|\l_{u,d}|^2 \kappa^* F \partial_{M^\ast} Z_{u,d},\qquad (m_{H_{u,d}}^2)_{LO} = -|\lambda_{u,d}|^2 |\kappa F|^2 \partial_M\partial_{M^*}Z_{u,d}
}}
with $Z_{u,d}\equiv \int d^4x\, \bra  \CO^\dagger_{u,d}(x) \CO_{u,d}(0)\ket_m$, where we have imagined deforming the messenger sector by $\delta W = M \CO_m$. This form of $A_{u,d}$ and $m^2_{H_{u,d}}$ indicates that, to leading order in ${F\over M^2}$, they generically arise at the same loop order in the messenger sector (when ``loop order'' is well-defined).   This is the GMHM analogue of the argument using field strength renormalization that was presented in \CraigXP. 

As shown in \CraigXP, MGM is the unique solution to the $A/m_H^2$ problem in the weakly-coupled messenger + spurion limit. In this case, the second derivative in \AmhProblemdoublet\ vanishes because the correlator in question evaluates to $\log M M^*$. One-loop contributions to $m_H^2$ still exist, but they are higher order in $\kappa$, i.e.\ they are suppressed by $F/M^2$. The same solution does not apply to $\mu/B_\mu$ because the relevant correlator does not generally factorize into terms holomorphic and anti-holomorphic in $M$, though it may be arranged in more elaborate models with additional scales \rattazi. In \CraigXP, the $\mu/B_\mu$ problem was avoided by taking $\lambda_d=0$, while $\mu$ and $B_\mu$ were then generated using an extension to the NMSSM along the lines of \refs{\GiudiceNI,\DelgadoRZ}.

Finally, let us comment on the ``little $A/m_H^2$ problem." This is manifested by the presence of the $A_{u,d}^2$ term in \softfinalspurion. Even if the 1-loop contribution to $m_{H_u}^2$ is dealt with through the MGM mechanism, large $A$-terms still imply a large 2-loop contribution to $m_{H_u}^2$, which drastically increases the tuning of the model or impedes electroweak symmetry breaking altogether. We emphasize that this is a universal feature of models with SUSY breaking spurions that generate large $A$-terms through Higgs-messenger couplings. The problem can ultimately be traced back to the relation $ \langle Q^4 \big[\CO_h(x)\CO_h^\dagger(0)\big]\rangle_{h,full} =|\langle Q^2 \CO_h\rangle_h|^2$, which is a consequence of the triviality of the spurion limit. When hidden sector interactions are accounted for, we may instead have $ \langle Q^4 \big[\CO_h(x)\CO_h^\dagger(0)\big]\rangle_{h,full} \ll |\langle Q^2 \CO_h\rangle_h|^2$, thus providing a route for solving the little $A/m_H^2$ problem. This strongly motivates going beyond the spurion limit in the hidden sector, as we consider in the next subsection. 

\subsec{Models with hidden sector SCFTs}

In these models we take $\sqrt{F}\ll M$, with the hidden sector described by an approximate SCFT at and above the scale $M$. Then, as discussed below \softparamgmgmdimtwofinal, the hidden sector correlator $\langle Q^4[\CO_h^\dagger(y) \CO_h(y')]\rangle_{h,full}$ is always pinned by the messenger sector correlator at $|y-y'|\lesssim {1\over M} \ll {1\over\sqrt{F}}$, i.e.\ at short distance. So we can apply the OPE of the SCFT to it:
\eqn\OhOPE{
\CO_h(y) \CO^\dagger_h(y')\sim  |y-y'|^{-2\D_h} {\bf 1} + \CC_\Delta |y-y'|^{\gamma}\CO_\D(y') + \dots
} 
where 
\eqn\gammadef{
\gamma\equiv \Delta-2\Delta_h
} 
Here ${\bf 1}$ is the unit operator (it drops out under the action of $Q^4$), and $\CO_\Delta$ (with dimension $\D$) is the lowest-dimension scalar operator in the UV fixed point of the hidden sector. The $\dots$  denotes terms with higher-dimension operators; we neglect them here as they will be further suppressed by $F/M^2$.  Substituting this into \softparamgmgmdimtwofinal\ we obtain
\eqn\softfinalCFT{\eqalign{
B_\mu &\approx \lambda_u\lambda_d |\kappa|^2\CC_\Delta \langle Q^4\CO_\Delta\rangle_h \int d^4y\,d^4y'\,  |y-y'|^{\gamma} 
  \Big\langle \CO_m(y)\CO_m^\dagger(y')\CX_{B_\mu}\Big\rangle_{m,full}\cr
m_{H_{u,d}}^2 &\approx -|\mu|^2 + |\lambda_{u,d}|^2  |\kappa|^2 \CC_\Delta\langle Q^4\CO_\Delta\rangle_h \int d^4y\,d^4y'\,   |y-y'|^{\gamma}  \Big\langle \CO_m(y)\CO_m^\dagger(y')\CX_{m_{H_{u,d}}^2}\Big\rangle_{m,full}
}}
As in the spurion limit, the general expressions for $\mu$ and $A_{u,d}$ again remain unchanged with respect to \softparamgmgm.
So if $\gamma>0$ (i.e.\ $\D>2\D_h$) and  $\sqrt{F}\ll M$, the contributions proportional to  $\langle Q^4\CO_\Delta\rangle_h$ are subleading with respect to those proportional to $|\bra Q^2 O_h\ket_h|^2$, and they are suppressed relative to $\mu^2$ and $A_{u,d}^2$. This is precisely the phenomenon of hidden-sector sequestering \refs{\DineDV\MurayamaGE-\schmaltz}, as seen from the point of view of GMHM. From \softfinalCFT, we note that the $-|\mu|^2$ contribution to $m_{H_{u,d}}^2$ is the only unsequestered contribution to the soft masses; in particular, there is no unsequestered contribution involving the OPE coefficient. We will comment more on the physical interpretation of this fact, and its relation to previous work, in the following section.

The idea of hidden-sector sequestering was originally proposed in order to solve the long-standing $\mu/B_\mu$ problem. Now with the need for large $A$-terms forced upon us by a Higgs at 125 GeV, we also have the $A/m_H^2$ problem to contend with. We see from \softfinalCFT\ that sequestering has the potential to solve both problems simultaneously. But despite its theoretical elegance, this approach suffers from a number of practical challenges. Foremost, it is difficult to achieve proper electroweak symmetry breaking with the fully sequestered boundary condition $B_\mu\approx0$ and $ m^2_{H_{u,d}}\approx -|\mu|^2$ \refs{\PerezNG, \AsanoQC}. 
Moreover, recent developments in the understanding of 4D SCFT's have resulted in strict upper bounds on the allowed anomalous dimensions \PolandEY. These bounds have made it increasingly difficult to envision a realistic setup where the anomalous dimensions and separation between $\sqrt{F}, M$ are large enough to achieve the desired amount of sequestering. 

The GMHM expressions \softfinalCFT\  point to possible ways out of these difficulties. For example, we see that the sequestered contributions in \softfinalCFT\ depend on the OPE coefficient $\CC_\Delta$. So if this is small for some reason, then we can again overcome the infamous loop factors. This is an entirely separate mechanism for solving the $\mu/B_\mu$ and $A/m_H^2$ problems that has not been considered before. Alternatively, one could combine a relatively small OPE coefficient with some realistic amount of sequestering. The expressions in \softfinalCFT\ provide a calculable setup to further investigate such partially sequestered models \futurework. 

In both of these solutions, the burden of addressing the $\mu/B_\mu$ and $A/m_H^2$ problems is shifted towards the hidden sector. This is in contrast to spurion-based models, where the messenger sector does all the legwork. From \softfinalCFT\ we can see another important difference with the spurion limit, as {\it both} the 1-loop and the 2-loop contributions are susceptible to sequestering and the smallness of the OPE coefficient. Therefore a solution to the $\mu/B_\mu$ and $A/m_H^2$ problems through sequestering, a small OPE coefficient, or some combination of the two, automatically implies a solution to the little $A/m_H^2$ problem.

\newsec{Comparison with effective theory}

Previous studies of hidden sector dynamics have worked in terms of the effective theory below the messenger scale $M$, in which the Higgs sector and hidden sector are coupled through irrelevant operators in the K\"{a}hler potential \refs{\DineDV\MurayamaGE\schmaltz\PerezNG- \CraigRK}. Furthermore, these studies have relied on using the RG to evolve down to the SUSY-breaking scale $\sqrt{F}\ll M$ in order to extract the physical soft parameters. In this section we re-visit the effective theory approach and show how its results can be matched to the GMHM calculation presented in Section 2 (which is more analogous to a fixed-order calculation in a full theory). 

The hidden sector may or may not be strongly coupled at the scale $M$. Either way, we will assume for simplicity that it is approximately superconformal, i.e.\ that $M$ is well-separated from all the other mass scales in the hidden sector. So we are in the SCFT limit of GMHM described in the previous subsection. Upon integrating out the messenger sector, the effective theory at the scale $M$ is of the form
\eqn\Keffops{\eqalign{ & \CL_{eff} \supset \int d^4 \theta \; \sum_i \Bigg[ {c_\mu(M) \over M^{\Delta_h}} \CO_h^\dag H_u H_d +  {c_{B_\mu,i}(M) \over M^{\Delta_i}} \CO_{\Delta_i} H_u H_d + {\rm h.c.}  \cr
&\qquad\qquad \qquad \qquad +{c_{A_{u,d}}(M) \over M^{\Delta_h}} \CO_h^\dag H_{u,d}^\dag H_{u,d} + {\rm h.c.} + {c_{m_{u,d},i}(M) \over M^{\Delta_i}} \CO_{\Delta_i} H_{u,d}^\dag H_{u,d}  \Bigg]
}}
where $\CO_h$ is a hidden-sector chiral operator with an $F$-term expectation value, while the $\CO_{\Delta_i}$ are non-chiral operators that appear in the OPE \OhOPE\ of $\CO_h^\dag$ and $\CO_h$. Previous approaches have only focused on the leading operator appearing in the OPE, but in general there are many such operators germane to the effective theory. For example, in Appendix B we construct an explicit Banks-Zaks example with two nontrivial $\CO_{\Delta_i}$. Unlike in the GMHM calculation, it will be important to keep track of all the operators in the OPE, because of the potentially unsequestered contributions in \hidsecseqclaimii.

The numerical values of the coefficients $c_\mu, c_{B_\mu,i}, c_{A_{u,d}}, c_{m_{u,d},i}$ at the scale $M$ depend on the details of the hidden sector and messenger sector, and they are fixed by matching to the full theory.  With this effective theory in hand, the Higgs sector soft parameters may be computed in three equivalent ways: (1) by direct calculation in the effective theory with cutoff $M$; (2) by RG evolution of the coefficients $c_i$ to a lower scale $E$ satisfying $\sqrt{F}\ll E < M$ followed by calculation in the effective theory (still assumed to be superconformal) with cutoff $E$; and (3) RG evolving down to a scale $E\ll \sqrt{F}$ and ``freezing-out" the SCFT dynamics by just substituting operator vevs, i.e.\ transitioning to the spurion limit where there are no nontrivial correlation functions. (Keep in mind that operator dimensions need not be canonical in the spurion limit.) The third  approach has been taken by previous works, with the further assumption that the transition to the spurion limit happens abruptly at $\sqrt{F}$. 
But it is very instructive to perform the calculation all three ways and compare with the predictions from GMHM. We can also compare the GMHM and direct effective theory results to arguments from field redefinitions when the hidden sector starts at a UV free fixed point. 

\subsec{Direct calculation in the effective theory with cutoff $M$}

To compute the Higgs sector soft parameters directly in the effective theory, we imagine performing the path integral over the effective theory with the momentum of hidden sector fields and loops of Higgs doublets restricted to lie below the Wilsonian cutoff $M$.  The leading contributions from operators involving the $\CO_{\Delta_i}$ are trivially computed by treating $H_{u,d}$ as background fields. There are also contributions to soft parameters quadratic in $c_{\mu}$ and $c_{A_{u,d}}$; these correspond to NLO contributions to the GMHM result coming from disconnected messenger correlators.

The calculation of the full soft mass proceeds entirely in parallel to the GMHM calculation. For simplicity we will focus on the scalar masses $m_{H_{u,d}}^2$; the calculation for $B_\mu$ is analogous. The linear $\CO_{\Delta_i}$ contributions are straightforward; working in the effective theory \Keffops\ to leading order in $c_{m_{u,d},i}$, we simply have:
\eqn\effectiveone{
m_{H_{u,d}}^2\Big|_{linear} = -\sum_i {c_{m_{u,d},i}(M)\over M^{\Delta_i}}\langle Q^4\CO_{\Delta_i}\rangle_M
}
Here and below, the subscript $M$ will denote correlation functions evaluated in the effective theory with cutoff $M$. Turning now to the contributions quadratic in $c_\mu$ and $c_{A_{u,d}}$, after some manipulations we have for example
 \eqn\effectivetwo{\eqalign{
m_{H_{u,d}}^2\Big|_{quadratic} \supset{ |c_{A_{u,d}}(M)|^2\over M^{2\Delta_h}} \int d^4x\, \left \langle Q^4 [\CO_h^\dag(x) \CO_h(0) ]\right \rangle_M \partial^2 \langle H_{u,d}^\dagger(x) H_{u,d}(0)\rangle_M
}}
At this stage, simply substituting a free propagator for the Higgs correlator in \effectivetwo\ is evidently problematic; integrating over $x$ would give rise to a pure contact term. This reflects the fact that the contribution being computed here is only sensitive to physics above the cutoff. Indeed, this agrees with the GMHM result -- as discussed at the end of Section 2, the hidden sector correlator for disconnected contributions is pinned by the messenger correlators at distances $\lesssim 1/M$, and so it does not  accumulate any significant contributions from below the scale $M$.

We can regulate the contact term in \effectivetwo\ any number of ways; different choices correspond to different prescriptions for matching with the full GMHM calculation. One useful regulator is to replace the delta function at $x=0$ with a (radial) delta function at $|x|=1/M$. As we will see below, this has the useful advantage of respecting both the physical cutoff at $M$ and the assumed abrupt transition to the spurion limit when the sliding cutoff is taken to $\sqrt{F}$. As such, we can apply this scheme uniformly to the various effective theory cases of interest and absorb all scheme-dependence into a single set of matching conditions at $M$. Substituting the general OPE \OhOPE\ and applying the regulator to \effectivetwo, we obtain
\eqn\effectivethree{\eqalign{
  m_{H_{u,d}}^2\Big|_{quadratic} &\supset \sum_i  {|c_{A_{u,d}}(M)|^2 \over M^{\Delta_i}}\CC_{\Delta_i} \langle Q^4 \CO_{\Delta_i} \rangle_M\cr
 }}
where $\CC_{\Delta_i}$ and $\gamma_i$ are defined as in \OhOPE. 
The calculation of the $|c_\mu|^2$ contribution is entirely analogous, although here we must remember to subtract out the fully disconnected contribution $|\mu|^2$, since this is conventionally not included in the definition of $m_{H_{u,d}}^2$. Repeating the same procedure for $B_\mu$ and combining the various contributions, the general effective theory result is 
\eqn\effectiveresult{\eqalign{
B_\mu &= -\sum_i \left({c_{B_\mu,i}(M) -  \CC_{\Delta_i} c_\mu(M) (c^\ast_{A_u}(M) + c^\ast_{A_d}(M) )  \over M^{\Delta_i}}   \right)\langle Q^4 \CO_{\Delta_i} \rangle_M\cr
m_{H_{u,d}}^2 &= -|\mu|^2 -\sum_i  \left(  {c_{m_{u,d},i}(M)- \CC_{\Delta_i}(|c_{A_{u,d}}(M)|^2 + |c_\mu(M)|^2) \over M^{\Delta_i}} \right)\langle Q^4 \CO_{\Delta_i} \rangle_M
}}
The dependence of the soft terms only on the composite vevs $\langle Q^4 \CO_{\Delta_i}\rangle$, and not on $|\langle Q^2 \CO_h \rangle|^2$, is in complete agreement with the result \softfinalCFT\ from GMHM in the SCFT limit. The specific linear combinations of coefficients appearing in \effectiveresult\ may be used to fix the matching of $c_{m_{u,d},i}, c_{B_\mu,i}$ with the $\CO(\lambda^4)$ terms in the GMHM result for $m_{H_{u,d}}^2$ and $B_\mu$.

\subsec{Effective theory with cutoff $\sqrt{F}\ll E<M$: testing the RGEs}
  
Alternately, we may compute the soft parameters in a different effective theory with a cutoff $E < M$ by evolving the coefficients $c_i(M)$ to the scale $E$ via RG running and repeating the direct calculation of scalar masses in the new effective theory. The scalar masses should, of course, agree with the result obtained in the theory with cutoff $M$. This procedure is completely straightforward in effective theories with cutoff $E \gg\sqrt{F}$, where the hidden sector is still an SCFT at the cutoff and the regularization scheme can be maintained. We refer the reader to \refs{\CraigRK, \GreenNQ} for the details of computing beta functions using superconformal perturbation theory. One thing to keep in mind is that to preserve the result for the soft masses, it is crucial to use the same regulator and scheme as in \effectivethree. The result for the beta functions between $M$ and $\sqrt{F}$ is:
\eqn\allbetas{\eqalign{
\beta_{c_\mu}&=\Delta_h c_\mu\cr
\beta_{c_{A_{u,d}}}&=\Delta_h c_{A_{u,d}}\cr
\beta_{c_{m_{u,d},i}} &= \Delta_i c_{m_{u,d},i} - \gamma_i  \CC_{\Delta_i}   (|c_\mu|^2 + |c_{A_{u,d}}|^2)\cr
\beta_{c_{B_\mu,i}} &= \Delta_i c_{B_\mu,i} - \gamma_i \CC_{\Delta_i} c_\mu ( c^\ast_{A_u} +c_\mu c^\ast_{A_d})
}}
In general, integrating the beta functions from $M$ to $E$ yields
\eqn\integratedbetas{\eqalign{
|c_\mu(E)|^2 &= |c_\mu(M)|^2 \left( {E \over M} \right)^{2 \Delta_h}  \; \; \; \; \; \; \; \; \;
|c_{A_{u,d}}(E)|^2 = |c_{A_{u,d}}(M)|^2 \left( {E\over M} \right)^{2 \Delta_h} \cr
c_{m_{u,d},i}(E) &= c_{m_{u,d},i}(M) \left( { E  \over M} \right)^{\Delta_i} -   \CC_{\Delta_i} \left( |c_\mu(M)|^2 + |c_{A_{u,d}}(M)|^2 \right) \left[ \left( {E \over M} \right)^{\Delta_i} - \left( {E \over M} \right)^{2 \Delta_h} \right]   \cr
c_{B_\mu,i}(E) &= c_{B_\mu,i}(M)\left( { E  \over M} \right)^{\Delta_i} -  \CC_{\Delta_i}   c_\mu(M) \left(c_{A_u}^\ast(M)+c_{A_d}^\ast(M) \right) \left[ \left( {E \over M} \right)^{\Delta_i} - \left( {E \over M} \right)^{2 \Delta_h} \right]
}}
The calculation of soft masses in the theory with cutoff $E$ proceeds in the SCFT limit precisely as above, with the replacement $M \to E$:
\eqn\effectiveCFTresulttwo{\eqalign{
B_\mu &= -\sum_i \left({c_{B_\mu,i}(E) -\CC_{\Delta_i} c_\mu(E) (c^\ast_{A_u}(E) + c^\ast_{A_d}(E) )\over E^{\Delta_i}} \right)\langle Q^4 \CO_{\Delta_i} \rangle_E\cr
m_{H_{u,d}}^2 &= -|\mu|^2 - \sum_i  \left(  {c_{m_{u,d},i}(E)- \CC_{\Delta_i}(|c_{A_{u,d}}(E)|^2 + |c_\mu(E)|^2) \over E^{\Delta_i}}   \right)\langle Q^4 \CO_{\Delta_i} \rangle_E
}}
Note that at a superconformal fixed point, operator wavefunction renormalization is trivial, and so $\langle Q^4 \CO_{\Delta_i} \rangle$ does not change between $M$ and $E$. Substituting the integrated couplings \integratedbetas\ into \effectiveCFTresulttwo, we again obtain \effectiveresult. Of course, the fact that running alone yields agreement between the two calculations is not surprising, since there are no physical thresholds between $M$ and $E$. This serves as an check of the RGEs that were derived independently using superconformal perturbation theory in \CraigRK.

 \subsec{Effective theory below $\sqrt{F}$}

Finally, we come to the most commonly considered case in the literature: RG evolving down to the scale $\sqrt{F}$ and ``freezing-out" the SCFT dynamics by just substituting operator vevs -- in other words, transitioning abruptly to the spurion limit. We imagine that just above the scale $\sqrt{F}$, some unspecified relevant operator in the SCFT turns on and drives it very quickly to a SUSY-breaking vacuum. Then right above $\sqrt{F}$, the couplings \integratedbetas\ obtained using the superconformal RGEs should be valid, while right below $\sqrt{F}$, the hidden sector is gapped and we should be in the spurion limit. Computing the dimension-two soft parameters in the spurion theory, we find:
\eqn\softparsrge{\eqalign{
 m_{H_{u,d}}^2 &= -|\mu|^2 - \sum_i {c_{m_{u,d},i}(\sqrt{F}) \over (\sqrt{F})^{\Delta_i} } \langle Q^4 \CO_{\Delta_i}\rangle_{\sqrt{F}} + {|c_{\mu}(\sqrt{F})|^2+|c_{A_{u,d}}(\sqrt{F})|^2 \over (\sqrt{F})^{2 \Delta_h}} |\langle Q^2 \CO_h \rangle_{\sqrt{F}} |^2  \cr
 B_\mu &=   -  \sum_i {c_{B_\mu,i}(\sqrt{F}) \over (\sqrt{F})^{\Delta_i} } \langle Q^4 \CO_{\Delta_i} \rangle_{\sqrt{F}}  + {c_\mu(\sqrt{F})(c^\ast_{A_u}(\sqrt{F})+c^\ast_{A_{d}}(\sqrt{F})) \over (\sqrt{F})^{2 \Delta_h}} |\langle Q^2 \CO_h \rangle_{\sqrt{F}} |^2 
}}
Substituting \integratedbetas\ into \softparsrge\ with $E\to \sqrt{F}$, we have
\eqn\softparsrgefinal{\eqalign{
m_{H_{u,d}}^2 &= -|\mu|^2- \sum_i \left( {c_{m_{u,d},i}(M) - \CC_{\Delta_i} \left( |c_\mu(M)|^2 + |c_{A_{u,d}}(M)|^2 \right) \over M^{\Delta_i} }  \right) \langle Q^4 \CO_{\Delta_i} \rangle_{\sqrt{F}}   \cr
&\qquad + {|c_\mu(M)|^2 + |c_{A_{u,d}}(M)|^2 \over M^{2 \Delta_h}}  \Delta S \cr
 B_\mu &= - \sum_i { c_{B_\mu,i}(M) -  \CC_{\Delta_i} c_\mu(M) \left(c_{A_u}^\ast(M)+c_{A_d}^\ast(M) \right) \over M^{\Delta_i} } \langle Q^4 \CO_{\Delta_i} \rangle_{\sqrt{F}}\cr
 & \qquad + {c_\mu(M)(c^\ast_{A_u}(M) + c^\ast_{A_d}(M) )\over M^{2 \Delta_h}}\Delta S
}}
where
\eqn\Deltadiff{
\Delta S \equiv  |\langle Q^2 \CO_h \rangle_{\sqrt{F}} |^2-\sum_i \CC_{\Delta_i} (\sqrt{F})^{-\gamma_i} \langle Q^4 \CO_{\Delta_i} \rangle_{\sqrt{F}}
 }
Comparing this with \effectiveresult, we see that there is an apparent disagreement. In particular, the answer in the spurion theory seems to have ``unsequestered" contributions $\propto ({\sqrt{F}/ M})^{2\Delta_h}$. This result illustrates the fact that, in general, threshold corrections  to the couplings $c_{B_\mu,i}$ and $c_{m_{u,d},i}$ at the scale $\sqrt{F}$ cannot be neglected. 

At the same time, it is also true that our regularization scheme (see the discussion around \effectivetwo) minimizes these threshold corrections. The key ingredient here is the continuity of the OPE. If the theory transitions abruptly to the spurion limit at a scale $\sqrt{F}$, then continuity of the OPE demands:
\eqn\vevsumrule{
|\langle Q^2\CO_h\rangle_{\sqrt{F}}|^2 \approx \langle Q^4[\CO_h(x)\CO_h(0)]\rangle_{\sqrt{F}}\big|_{|x|=1/\sqrt{F}} \approx \sum_i \CC_{\Delta_i} (\sqrt{F})^{-\gamma_i} \langle Q^4 \CO_{\Delta_i} \rangle_{\sqrt{F}} 
}
Substituting this into \softparsrgefinal, we find that the threshold corrections are minimized, and the result is brought into agreement with previous effective theory calculations and the GMHM expectation. In the limit $\Delta_i \gg 2 \Delta_h$, the $\langle Q^4 \CO_{\Delta_i} \rangle$ terms are negligible, and we indeed find 
\eqn\fullseqindeed{
m_{H_{u,d}}^2 \approx -|\mu|^2, \qquad B_\mu \approx 0
}
as claimed in \refs{\MurayamaGE,\PerezNG}. 

Had we chosen a different regularization scheme in \effectivetwo, e.g.\ a smoother regulator that smears out the integrand in \effectivetwo\ around the cutoff, then the second term in $\Delta S$ would have been correspondingly smeared. Then the OPE sum rule \vevsumrule\ would not have accounted for $\Delta S$, and additional threshold corrections to the couplings would have been required.\foot{Even in this case, one can check that in the limit $\gamma_i \ll 1$, the beta functions and soft masses become scheme independent to leading order in $\gamma_i$ and the matching procedure at $\sqrt{F}$ is likewise insensitive to the details of exiting the SCFT.}
 In a sense, our choice of radial delta function regularization is particularly appropriate in that it is abrupt and localized at the cutoff, in the same way that our transition to the spurion limit is taken to be abrupt. This allows us to maintain the same regularization scheme in effective theories above and below $\sqrt{F}$ and smoothly absorb all scheme dependence into matching at the scale $M$.

\subsec{Field redefinitions}

Finally, we can compare both results to the soft parameter predictions obtained via field redefinition at a UV free fixed point as in \PerezNG. If the hidden sector is weakly interacting at the scale $M$, and $\CO_h$ is a dimension-one elementary field, then the only nontrivial operator in the OPE is $\CO_{\Delta_1} \sim \CO_h^\dag \CO_h$ and the terms linear in $\CO_h$ in \Keffops\ are redundant, i.e., may be eliminated by field redefinitions.  Such UV free theories are only a restricted subset of models amenable to treatment by our formalism, but they provide a useful check.

 In this case, the terms proportional to $c_{A_{u,d}}$ may be eliminated by the redefinition
\eqn\Aredef{
H_{u,d} \to \tilde H_{u,d} = H_{u,d} (1 + c_{A_{u,d}} \CO^\dag_h /M)
}
which leads to an equivalent effective theory at the scale $M$
\eqn\Keffredefone{\eqalign{\CL \supset \int d^4 \theta \; \left[ {c_\mu \over M} \CO_h^\dag \tilde H_u \tilde H_d + {c_{B_\mu} - c_{\mu}(c^\ast_{A_u} + c^\ast_{A_d}) \over M^2} \CO_h^\dag \CO_h \tilde H_u \tilde H_d + {\rm h.c.} \right. \cr
\left. + {c_{m_{u,d}} - |c_{A_{u,d}}|^2 \over M^2} \CO_h^\dag \CO_h \tilde H_{u,d}^\dag \tilde H_{u,d} + \dots  \right]
}}
where the ellipses denote terms of cubic order or higher in hidden sector fields (i.e., higher order in $\kappa$ in the GMHM approach). In this effective theory there are no additional contributions to Higgs soft masses proportional to $|c_{A_{u,d}}|^2$.

To compute the physical mass of the scalar doublet $\tilde H_{u}$, we may treat it as a background field, keeping $\tilde H_d, \CO_h$ as dynamical fields and performing the field redefinition 
\eqn\muredef{
\tilde H_{d} \to \tilde H_d' = \tilde H_d + {c_\mu^* \over M} \CO_h \tilde H_u^\dag
}
where the apparently non-holomorphic field redefinition preserves supersymmetry because $\tilde H_u$ is simply a background field. Now there are also no additional contributions proportional to $|c_\mu|^2$, and the calculation of soft masses is straightforward. In this theory the physical mass of the scalar $\tilde H_u$ is simply
\eqn\humass{
 m_{\tilde H_{u}}^2 = -|\mu|^2 - {c_{m_u}(M) - |c_{A_u}(M)|^2 - |c_\mu(M)|^2 \over M^2} \langle Q^4 ( \CO_h^\dag \CO_h) \rangle_M
}
From this we can infer the soft mass, and it is in complete agreement with the results from GMHM and the effective theory. Analogous arguments hold for the calculation of $m_{H_d}^2$ and $B_\mu$. Note that here $\langle Q^4 ( \CO_h^\dag \CO_h) \rangle_M \neq |F|^2$, since by assumption the hidden sector is asymptotically free and the operator vev reflects sequestering due to nontrivial dynamics below the scale $M$.

To summarize, we have found agreement between the Higgs sector soft parameters as computed in GMHM and the soft parameters computed by a variety of approaches in the effective theory below the messenger scale: directly in the effective theory defined at the scale $M$; in effective theories with cutoffs above and below $\sqrt{F}$; and via field redefinition in the effective theory when the hidden sector is weakly coupled at the scale $M$. The key to reconciling the weakly-coupled results of \refs{\MurayamaGE,\PerezNG} with the superconformal perturbation theory result of \CraigRK\ is the approximate operator vev sum rule \vevsumrule\ imposed by the OPE. 

Of course, thus far our discussion has remained fairly abstract. We validate certain features of our analysis by comparison with explicit perturbative calculations in a toy Banks-Zaks model, the details of which we reserve for Appendix B.

\newsec{Conclusions and future directions}

The discovery of a Higgs near 125 GeV poses significant challenges for minimal supersymmetry. If electroweak symmetry breaking is natural, either the Higgs sector must be extended -- often at the expense of other attractive features of the MSSM such as perturbative gauge coupling unification -- or $A$-terms must be large. While this latter option is attractive, it poses a particular challenge for calculable models where intrinsic $A$ terms are naturally small. Introducing new interactions to generate $A$ terms results in the $A/m_H^2$ problem, i.e., unwanted contributions to other soft terms that threaten EWSB and supersymmetric naturalness.

Yet the $A/m_H^2$ problem is but one symptom of a broader sickness in the Higgs sector of calculable models. Beyond confronting the $A/m_H^2$ problem to accommodate the observed Higgs mass, calculable models must also confront the more familiar $\mu/B_\mu$ problem to achieve EWSB. In addition, we see from \dimtwofull\ that such models also potentially suffer from the little $A/m^2_H$ problem, even if the one loop contribution to $m^2_{H_{u}}$ vanishes.  The ubiquity and tenaciousness of these problems in calculable models with weakly-coupled hidden sectors strongly favors hidden sectors with non-trivial dynamics. In this case, powerful tools are required in order to make concrete predictions for the physical spectrum.

In this work, we have developed a framework for computing the soft spectrum arising from general Higgs-messenger interactions in theories where the SUSY-breaking dynamics factorizes into arbitrary messenger and hidden sectors. We compute soft parameters in a supersymmetric correlator formalism   through a double expansion in the portals connecting the Higgs, messenger, and hidden sectors. This approach allows us to identify general solutions to the $\mu/B_\mu$ and $A/m_H^2$ problems. An essential key is that while $\mu$ and $A_{u,d}$ depend on the one-point function $\langle Q^2 \CO_h\rangle_h$ in the hidden sector, $B_\mu$ and $m_{H_{u,d}}^2$ depend on the two-point function $\langle Q^4 (\CO_h^\dag(y) \CO_h(y')) \rangle_h$. Although in spurion models these two are trivially related, more generally they need not have anything to do with one another.

Although our results are quite general, we demonstrate their power by using them to compute the soft spectrum for hidden sectors in the spurion and SCFT limits. In the SCFT limit, we make contact with previous approaches to hidden sector sequestering \refs{\DineDV\MurayamaGE \schmaltz\PerezNG-\CraigRK}. In particular, we resolve a long-standing disagreement between different approaches to hidden-sector sequestering, validating the results obtained via field redefinitions and reconciling previously conflicting results from superconformal perturbation theory using an approximate sum rule derived from the OPE. However, our general formalism allows us to go beyond the case of full sequestering considered in previous works and compute the soft spectrum in the case of partial sequestering, where hidden sector anomalous dimensions conspire with details of the hidden sector to yield potentially viable phenomenology. This is particularly attractive since the idealized limit of full sequestering appears increasingly unrealistic due to both limits on operator dimensions \PolandEY\ and tightly constrained parametrics \refs{\PerezNG,\AsanoQC}. In partially sequestered scenarios, SCFT data (such as OPE coefficients and operator dimensions), operator vevs, and numerical coefficients all play important roles in solving the $\mu/B_\mu$ and $A/m_H^2$ problems. Interestingly, in contrast with the spurion limit, a solution to the $A/m_H^2$ problem in this context automatically guarantees a solution to the little $A/m_H^2$ problem. Moreover these models have much more parametric freedom compared to the fully sequestered case, and exhibit novel phenomenology that we will explore in detail in future work \futurework.
 
Let us conclude by highlighting a variety of interesting future directions:

\lfm{1.} Much as GGM delineated the full parametric freedom available in gauge mediation, our formalism delineates the full parametric freedom available to models with Higgs-messenger interactions. It would be particularly useful to determine whether this full parameter space may be spanned by weakly coupled models, along the lines of what was done for GGM in  \refs{\BuicanWS,\CarpenterWI}.

\lfm{2.} In this work we have applied our formalism to two simplified cases, the spurion limit and the SCFT limit. However, the formalism may be applied to any theory in which the overall hidden sector factorizes into separate messenger and SUSY-breaking hidden sectors, and there are likely many other well-motivated cases amenable to detailed study. For example, it may be used to compute corrections to the spurion limit in weakly-interacting hidden sectors whose IR physics are described by O'Raifeartaigh models. 

\lfm{3.} We have restricted our attention to superpotential portals connecting the messenger and hidden sectors. It would be interesting to analyze K\"{a}hler portals to determine whether there are other qualitatively new features or new approaches to the $\mu/B_\mu$ and $A/m_H^2$ problems. Along similar lines, we have focused on SUSY-breaking due to a chiral operator in the hidden sector; it would be interesting to consider more general operators as well.

\lfm{4.} Considerable attention has recently been devoted to the NMSSM in light of the observed Higgs mass, and calculable models for the NMSSM soft spectrum must confront challenges analogous to the $A/m_H^2$ problem. It would therefore be fruitful to extend our formalism to cover the NMSSM and related models involving additional degrees of freedom at the weak scale.

\lfm{5.} In partially sequestered scenarios, SCFT data such as operator dimensions and OPE coefficients play a key role in determining the Higgs soft spectrum.  While considerable effort has recently been devoted to developing general bounds on operator dimensions in 4D SCFTs \PolandEY, it would be particularly useful to extend general bounds on OPE coefficients beyond those considered in \PolandEY. This in turn should increase the predictiveness of viable partially-sequestered models. 

\break

\noindent {\bf Acknowledgments:}

We would like to thank T.~Banks, D.~Green, M.~Schmaltz, N.~Seiberg, S.~Thomas, and Y.~Zhao  for
useful discussions. NC is supported by NSF grant PHY-0907744, DOE grant DE-FG02-96ER40959, and the Institute for Advanced Study. SK is supported by DOE grant DE-FG02-96ER40959. DS is supported in part by a DOE Early Career Award and a Sloan Foundation Fellowship.

\appendix{A}{Details on the factorization of the correlators}

In this appendix, we provide some more details about the required $R$-symmetry in the messenger sector and show explicitly how the correlators for the dimension-two soft parameters in \softparamgmgm\ factorize into a hidden sector correlator and a messenger sector correlator. 

\subsec{$\CO(\kappa)$ expansion and $R$-symmetry}
With the $\CX$ defined in \Xdef, the formulas for the soft parameters to leading order in $\kappa$ are given by
\eqn\softparamorderkappa{\eqalign{
  \mu =& \lambda_u\lambda_d \kappa^* \langle \Q^2 \CO_h^\dagger \rangle_h\int d^4 y\; \langle \CO^\dagger_m(y) \CX_\mu\rangle_m\
  + \lambda_u\lambda_d\kappa\langle Q^2 \CO_h \rangle_h\int  d^4 y\; \langle \CO_m(y) \CX_\mu\rangle_m\cr
  A_{u,d} =& |\lambda_{u,d}|^2\kappa^* \langle \Q^2 \CO_h^\dagger \rangle\int  d^4 y\; \langle \CO^\dagger_m(y)\CX_{A_{u,d}}\rangle_m\cr
  \hat B_\mu =&\lambda_u\lambda_d\kappa \langle Q^2 \CO_h\rangle_h \int  d^4 y\; \langle \CO_m(y) \CX_{B_\mu}\rangle_m\cr
  \hat m_{H_{u,d}}^2 =&\CO(\kappa^2)
 }}
All other possibilities are forbidden by the supersymmetry Ward identities on the messenger correlator. This can be most easily understood from the observation that $\CX_{A_{u,d}}$, $\CX_{B_\mu}$ and $\CX_{ m_{H_{u,d}}^2}$ in \Xdef\ can be written as 
\eqn\Xrewrite{\eqalign{
\CX_{A_{u,d}}=\Q^2 \big[\cdots \big],\quad
\CX_{B_\mu}=Q^2 \big[\cdots\big],\quad
\CX_{ m_{H_{u,d}}^2}=Q^4\big[\cdots\big]
}}
where the $\cdots$ are integrated operators built out of supercharges, $O_{u,d}$ and $H_{u,d}$ operators. To understand the formula for $\mu$, one should additionally keep in mind that we have shifted the vev of the lowest component of $\CO_h$ to zero, as explained in the introduction. 

The $\CO(\kappa)$ contribution to $B_\mu$ in \softparamorderkappa\ is allowed by supersymmetry and results in the parametric behavior $B_\mu \sim M \mu$, which is disastrous for electroweak symmetry breaking. We therefore wish to impose a suitable symmetry on the messenger sector that forbids the correlator contributing to $B_\mu$, while preserving an $\CO (\kappa)$ contribution to both $\mu$ and $A_{u,d}$. From \softparamorderkappa\ one can easily see that the only symmetries satisfying these criteria are $R$-symmetries with charge assignments 
\eqn\Rassignone{
R[\CO_m]=2 \quad R[\CO_u]+R[\CO_d]=4
}
or 
\eqn\Rassigntwo{
R[\CO_m]=2 \quad R[\CO_u]+R[\CO_d]=0
}
\Rassignone\ and \Rassigntwo\ respectively preserve the first and second correlator contributing to $\mu$. All known models in the literature adhere to the first charge assignment, and this is why we have assumed \Rassignone\ throughout this paper. It would of course be interesting to explore the other $R$-charge assignment, but we will not do so here. We emphasize that the presence of an $R$-symmetry in the messenger sector is a generic feature of all models that attempt to generate both $\mu$ and $A_{u,d}$ through the same set of Higgs-Messenger interactions. 

\subsec{$\CO(\kappa^2)$ expansion}

Since the $\CO(\kappa)$ contribution is assumed to vanish by virtue of the $R$-symmetry that we imposed in the previous section, we now proceed to the derivation for the $\CO(\kappa^2)$ contribution to $\hat B_\mu$. The derivation for $\hat m^2_{H_{u,d}}$ is completely analogous.  The only contribution compatible with the $R$-symmetry in \Rassignone\ is
\eqn\BmuAppendix{\eqalign{
& \hat B_\mu = \lambda_u\lambda_d|\kappa|^2 \int d^4y\,d^4y'\,\Big \langle Q^2\big[\CO_h\CO_m(y)\big] \Q^2\big[\CO^\dagger_h\CO^\dagger_m(y')\big]   \CX_{B_\mu} \Big\rangle_{m+h} 
}}
Using \Xrewrite, we can write \BmuAppendix\ as
\eqn\BmuAppendixTwo{\eqalign{
& \hat B_\mu = \lambda_u\lambda_d |\kappa|^2 \int d^4y\,d^4y'\,\Big \langle Q^2\Q^2\Big[\CO_h\CO_m(y) \CO^\dagger_h\CO^\dagger_m(y')  \CX_{B_\mu} \Big] \Big\rangle_{m+h} 
}}
where we dropped total derivatives. Now we redistribute the supercharges over the combinations $\CO_h\CO_h^\dagger$ and $\CO_m\CO^\dagger_m  \CX_{B_\mu}$ and factorize the correlators.  The unbroken supersymmetry of the messenger correlator kills all terms except the term where all the supercharges are inside the hidden sector correlator:
\eqn\BmuAppendixThree{\eqalign{
& \hat B_\mu = \lambda_u\lambda_d|\kappa|^2 \int d^4y\,d^4y'\,\Big \langle Q^2\Q^2\Big[\CO_h(y)\CO_h^\dagger(y')\Big]\Big \rangle_h\Big \langle \CO_m(y)\CO^\dagger_m(y')  \CX_{B_\mu}\Big\rangle_{m} 
}}
The result has now arranged itself such that all the contributions from the hidden sector are packaged in a single hidden sector two-point function.

\subsec{Short distance dominance of the messenger correlator}

Finally, let us explicitly verify that the disconnected components of the messenger correlators for $B_\mu$ and $m_{H_{u,d}}^2$ indeed fall off at long distance as claimed in \falloff, and that they integrate to give the auxiliary field contributions in \dimtwofull, as claimed in \softparamgmgmdimtwofinal.

As an example, consider the $\CO(\kappa^2, \lambda_u\lambda_d|\lambda_u|^2)$ messenger correlator for $B_\mu$
\eqn\corfactorize{\eqalign{
& \Big \langle \CO_m(y)\CO^\dagger_m(y')  \CX_{B_\mu}\Big\rangle_{m} \cr
 &\quad \supset |\lambda_u|^2 \int d^4x d^4z d^4z'\, \langle \CO_m(y) \CO_m^\dagger (y') Q^2 \CO_u(x) Q^2 \CO_d(0) Q^2\big[\CO_uH_u\big](z)\Q^2\big[\CO^\dagger_uH^\dagger_u\big](z')\rangle_m\cr
&\quad\,\, \,\,= |\lambda_u|^2 \int d^4x d^4z d^4z'\, \langle Q^2 \CO_m(y) \CO_m^\dagger (y') Q^2 \CO_u(x) Q^2 \CO_d(0) \CO_uH_u(z)\Q^2\big[\CO^\dagger_uH^\dagger_u\big](z')\rangle_m
}}
where in the second line we used the supersymmetry Ward identity, dropping any total derivatives. In order to enable a contraction between the $H_u$ operators, the $\Q^2$ must act on $O_u^\dagger(z')$:
\eqn\corfactorizei{\eqalign{
& \Big \langle \CO_m(y)\CO^\dagger_m(y')  \CX_{B_\mu}\Big\rangle_{m}\cr
 & \quad \supset |\lambda_u|^2 \int d^4x d^4z d^4z'\, {1\over 4\pi^2}{1\over (z-z')^2}\langle Q^2 \CO_m(y) \CO_m^\dagger (y') Q^2 \CO_u(x) Q^2 \CO_d(0) \CO_u(z)\Q^2\CO^\dagger_u(z')\rangle_m
}}
Now we want to factorize this into two separate correlators. All one-point functions are assumed to vanish, and one can easily check that there is no factorization into a product of two- and four-point functions consistent with the symmetries. This leaves a product of three-point functions, and here the only non-vanishing factorization
consistent with all the symmetries is 
\eqn\corfactorizeii{\eqalign{
& \Big \langle \CO_m(y)\CO^\dagger_m(y')  \CX_{B_\mu}\Big\rangle_{m}\cr
 & \quad \supset |\lambda_u|^2 \int d^4x d^4z d^4z'\, {1\over 4\pi^2}{1\over (z-z')^2}\langle Q^2 \CO_m(y) Q^2 \CO_u(x) \Q^2\CO^\dagger_u(z') \rangle_m \langle  \CO_m^\dagger (y')  Q^2 \CO_d(0) \CO_u(z)\rangle_m\cr
}}
Moving the supercharges around in the first correlator produces a $\partial_{z'}^2$, and after integrating by parts we are left with a $\delta^{(4)}(z-z')$. So the answer becomes:
\eqn\corfactorizeiii{\eqalign{
& \Big \langle \CO_m(y)\CO^\dagger_m(y')  \CX_{B_\mu}\Big\rangle_{m}\cr
 & \quad \supset |\lambda_u|^2 \int d^4x d^4z \,\langle  \CO_m(y) Q^2 \CO_u(x) \CO^\dagger_u(z) \rangle_m \langle  \CO_m^\dagger (y')  Q^2 \CO_d(0) \CO_u(z)\rangle_m
}}
This is the desired result: the delta function ensures that, to this order in perturbation theory, the messenger correlator always falls off exponentially for $M|y-y'|\to\infty$, despite the fact that it is disconnected. Furthermore, substituting \corfactorizeiii\ back into \BmuAppendixThree\ but now using the disconnected component of the hidden sector correlator, the result becomes precisely $\mu A_u^*$. The argument is analoguous for $m^2_{H_{u,d}}$ and the $\CO(\kappa^2, \lambda_u\lambda_d|\lambda_d|^2)$ contribution to $B_\mu$, and in this way we reproduce the contributions from integrating out the $F$-terms in \dimtwofull.

\appendix{B}{A Banks-Zaks model of hidden sector renormalization }

\subsec{Setup}

In this appendix we study a toy example of a weakly-coupled interacting SCFT containing a chiral gauge singlet operator $X$ that will serve as a proxy for the supersymmetry breaking operator $\CO_h$. The goal is to validate the beta functions in \allbetas, obtained through superconformal perturbation theory, against a direct calculation of the beta functions through Feynman diagrams. In the process, we are also equipped to confirm the validity of the field redefinition argument in our weakly coupled example. This provides an explicit check of the various approaches to hidden sector sequestering in an effective theory framework. 

Our toy model is the same one as in \GreenNQ: a Banks-Zaks model coupled to $X$ via the superpotential
\eqn\Wdef{
W ={1\over2\pi} \lambda X {\rm Tr}\,Q \tilde Q ~.
}
Here $Q$ and $\tilde Q$ are $N_f$ flavors charged under an $SU(N_c)$ gauge group with $N_f = 3 N_c / (1+\epsilon)$, and the trace is over all colors and flavors.\foot{In what follows, we work with the conventions in \GreenNQ. In particular, we take $Q$ and $\tilde Q$ to be canonically normalized, and $X$ to be CFT-canonically normalized.} Before the deformation \Wdef, the only coupling in the theory is the gauge coupling, $g$; for $N_f, N_c \gg 1$,
this undeformed theory flows to the perturbative BZ fixed point at which $\beta_{g} = 0$. Deforming this BZ model by the addition of \Wdef\ induces a flow to a new fixed point at which $\beta_{\lambda} = \beta_{g} = 0$, with \GreenNQ
\eqn\fpcplgl{\eqalign{
& \hat g_{*} = \left(\epsilon + \CO(\epsilon^2) \right)  + {5\over 3N_c^2}\Big(\epsilon+\CO(\epsilon^2)\Big) \cr
& \hat\lambda_* = {2\over3}\epsilon(1+\epsilon)+ \CO\left(\epsilon^2\over N_c^2\right) 
}}
where $\hat{g} = N_c g^2 / 8 \pi^2,  \hat\lambda = N_c^2 \lambda^2 / 8\pi^2$ are the generalized 't Hooft couplings.

We would like to study hidden sector renormalization at the fixed point in which K\"{a}hler operators linear in $X$ renormalize K\"{a}hler operators linear in $\CO_{\Delta_i}$, where $\CO_{\Delta_i}$ are scaling operators that appear in the OPE of $X^\dagger X$. In a perturbative SCFT, we expect the dimensions of such operators $\CO_{\Delta_i}$  to be close to two. This singlet-deformed BZ theory possesses two such candidate operators:
\eqn\LJXdef{\eqalign{
L &= {4\pi^2\over \sqrt{2N_fN_c}} {\rm Tr}(Q^\dagger Q+\tilde Q^\dagger \tilde Q)\cr
J^X &= X^\dagger X ~.
}}
Here these operators are CFT-canonically normalized to leading order in the undeformed theory, i.e., using free field contractions. These operators are easiest to work with from the point of view of computing Feynman diagrams. But due to mixing in the beta functions, they are not scaling operators at the deformed BZ fixed point. Rather, they are related to  scaling operators $\CO_{\Delta_1}$ and $\CO_{\Delta_2}$ via a linear transformation:
\eqn\LJXscaling{\eqalign{
 \pmatrix{ \CO_{\Delta_1}\cr \CO_{\Delta_2}} = \pmatrix{ S_{11} & S_{12} \cr S_{21} & S_{22}} \pmatrix{ L \cr J^X} \ .
}} 
This change of basis is related to the diagonalization of the matrix of anomalous dimensions $\Gamma$; we refer the reader to \GreenNQ\ for details. Although explicit formulas for $S$ can be derived, we will not actually need them.

\subsec{Beta functions}

Here we will verify the renormalization of the couplings $c_{B_\mu,i}$ and $c_{m_{u,d},i}$ due to $c_\mu, c_{A_{u,d}}$. To do so, we compute this result at the fixed point using superconformal perturbation theory and compare with data computed perturbatively at the free fixed point.

Neglecting visible-sector interactions due to $H_u, H_d$, for the couplings $c_{m_{u,d},i}, c_{B_\mu,i}$ we apply the superconformal perturbation theory results of Section 4 to find
\eqn\betagen{\eqalign{
& {d\over d\log\Lambda} \pmatrix{ c_{m_{u,d},1}\cr c_{m_{u,d},2}} = \pmatrix{ \Delta_1 & 0 \cr 0 & \Delta_2}\pmatrix{ c_{m_{u,d},1}\cr c_{m_{u,d},2}} - \pmatrix{ \CC_{\Delta_1}\gamma_1(1+\dots)\cr \CC_{\Delta_2}\gamma_2(1+\dots)}  (|c_\mu|^2 + |c_{A_{u,d}}|^2) \cr
& {d\over d\log\Lambda} \pmatrix{ c_{B_\mu,1}\cr c_{B_\mu,2}} = \pmatrix{ \Delta_1 & 0 \cr 0 & \Delta_2}\pmatrix{ c_{B_\mu,1}\cr c_{B_\mu,2}} - \pmatrix{ \CC_{\Delta_1}\gamma_1(1+\dots)\cr \CC_{\Delta_2}\gamma_2(1+\dots)} (c_\mu c^\ast_{A_u} + c_\mu c^\ast_{A_d})~.
}}
where again $\gamma_i \equiv \Delta_i - 2 \Delta_X$. The $\dots$ are higher order corrections in $\gamma_{1,2}$ that are scheme-dependent. For comparison with the direct calculation of RGEs, we need to compute the coefficients $\CC_{\Delta_1} \gamma_1$ and $\CC_{\Delta_2} \gamma_2$. While we could compute the OPE coefficients and anomalous dimensions separately, it suffices to merely extract the combinations $\CC_{\Delta_i} \gamma_i$ from three-point functions in the hidden sector. The form of the three-point functions is dictated by conformal invariance up to the OPE coefficients $\CC_{\Delta_1}$, $C_{\Delta_2}$; expanding in powers of $\epsilon$ and $1/N_c$ yields
\eqn\threeptfnexpand{\eqalign{
& \langle X^\dagger(x_1) X(x_2) \CO_1(x_3)\rangle = {\CC_{\Delta_1}\over x_{13}^2x_{23}^2} \left( 1 +\gamma_1\log x_{12} - \nu_1 \log (x_{13}x_{23})\right) + \dots \cr
& \langle X^\dagger(x_1) X(x_2) \CO_2(x_3)\rangle =  {\CC_{\Delta_2}\over x_{13}^2x_{23}^2} \left( 1 +\gamma_2\log x_{12} - \nu_2 \log (x_{13}x_{23})\right) + \dots
}}

Working around the free fixed point, we have access to the $X^\dagger-X-L$ and the $X^\dagger-X-J^X$ three point functions, shown diagrammatically at  one loop in fig.\ 5. We can compute $\CC_{\Delta_i} \gamma_i$ by isolating the $\log x_{12}$ terms in the perturbative three-point functions and rotating to the basis of scaling operators using \LJXscaling.

\ifig\opediagrams{The leading perturbative contributions to the $X^\dagger - X - L$ and $X^\dagger-X-J^X$ three point functions. }{\epsfxsize=3cm\epsfbox{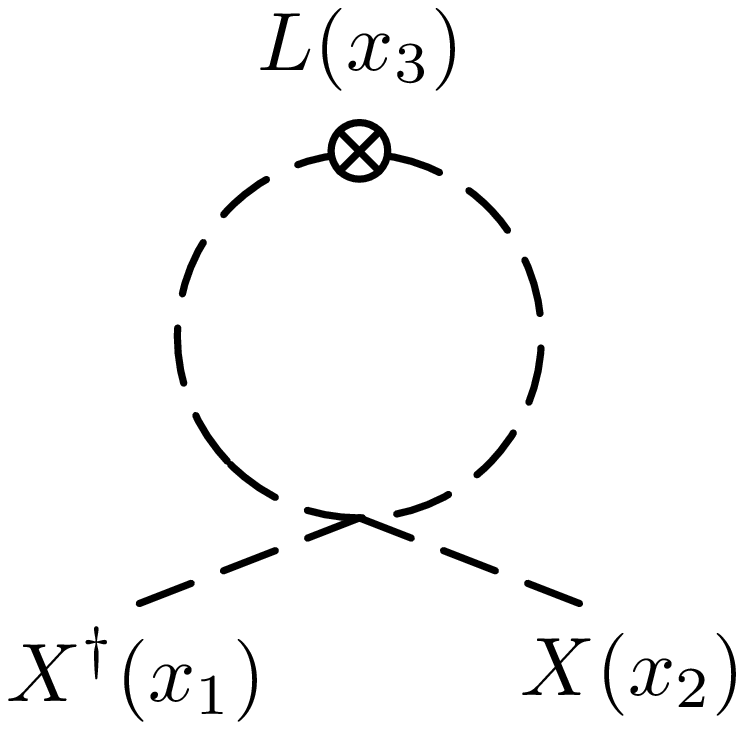} \epsfxsize=6cm\epsfbox{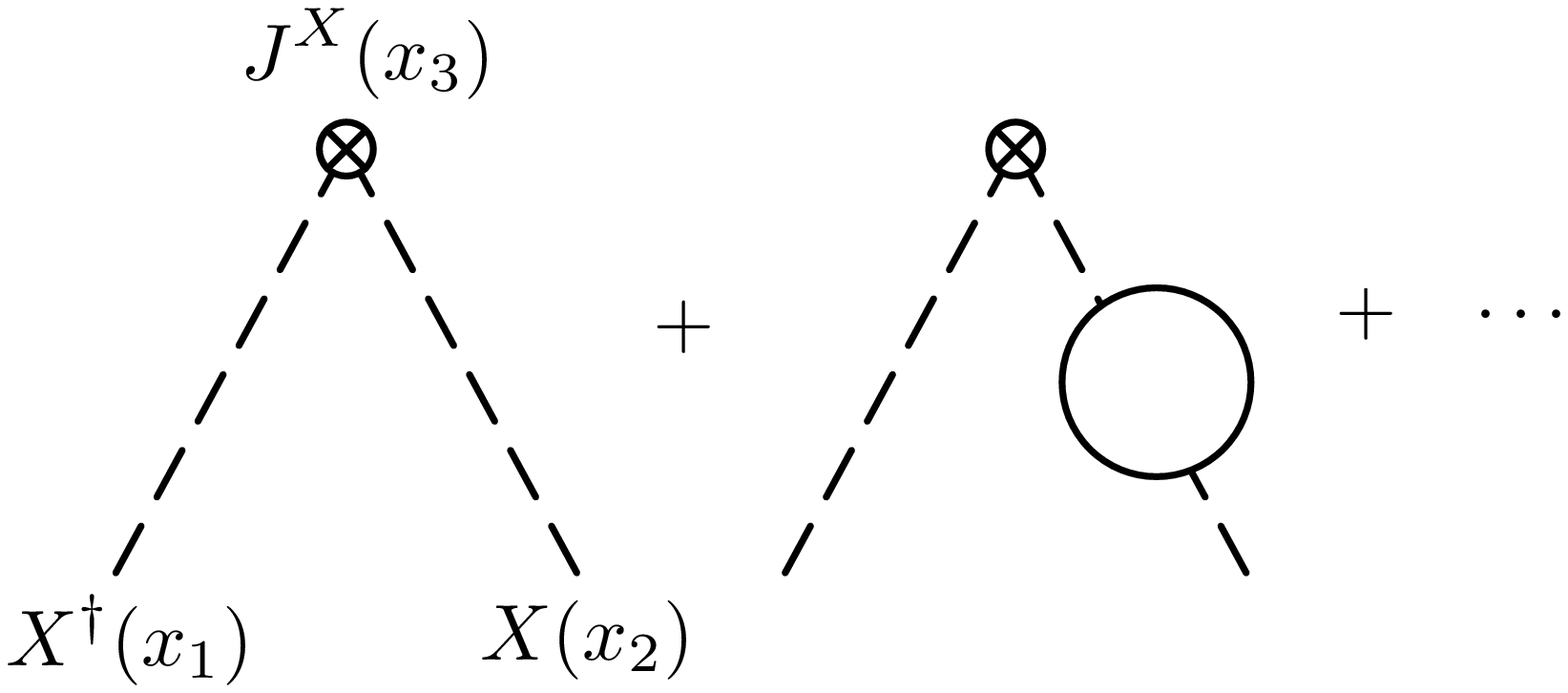}} 

Therefore we have at one loop
\eqn\findCg{\eqalign{
 \pmatrix{ \CC_{\Delta_1} \gamma_1 \cr \CC_{\Delta_2} \gamma_2} =  S \pmatrix{ b_L \cr b_{J^X} }
 }} 
 where $b_L, b_{J^X}$ are the coefficients of the $\log x_{12}$ terms appearing in the $X^\dagger-X-L$ and $X^\dagger-X-J^X$ three point functions, respectively. Diagrammatically, it is clear that $b_{J^X} = 0$ at one loop, since the loops in $X^\dagger-X-J^X$ are functions only of $x_{13}$ or $x_{23}$. However, the loop in $X^\dagger-X-L$ is sensitive to $x_{12}$, and so $b_L$ should be nonzero at one loop. An explicit calculation of the diagrams in \opediagrams\  yields $b_L = 2 \sqrt{2 \over 3} {\epsilon \over N_c}$ and $b_{J^X} = 0$.
 
 Now we can compare the superconformal perturbation theory result with standard pertubation theory around the free fixed point. As before, the calculation around the free fixed point in terms of $L, J^X$ is related to the scaling operators by the transformation \LJXscaling. Thus we need only verify that 
 \eqn\betagenii{\eqalign{
 {d\over d\log\Lambda} \pmatrix{ c_{m_{u,d},L}\cr c_{m_{u,d},J^X}} 
&\supset 
 - \pmatrix{ b_L \cr b_{J^X}} (|c_\mu|^2 +|c_{A_{u,d}}|^2) \cr
  {d\over d\log\Lambda} \pmatrix{ c_{B_\mu,L}\cr c_{B_\mu,J^X}} 
&\supset
 - \pmatrix{ b_L  \cr b_{J^X}} (c_\mu c_{A_u}^\ast + c_\mu c_{A_d}^\ast)
}}
 by a standard one-loop calculation of beta functions around the free fixed point.
 
\ifig\bzdiagrams{Component diagrams for the one-loop renormalization of (a) $c_{m_u,L}$ proportional to $|c_\mu|^2$, (b) $c_{m_u,L}$ proportional to $|c_{A_u}|^2$, and (c) $c_{B_\mu,L}$ proportional to $c_\mu c_{A_u}^\ast$. }{\epsfxsize=0.7\hsize\epsfbox{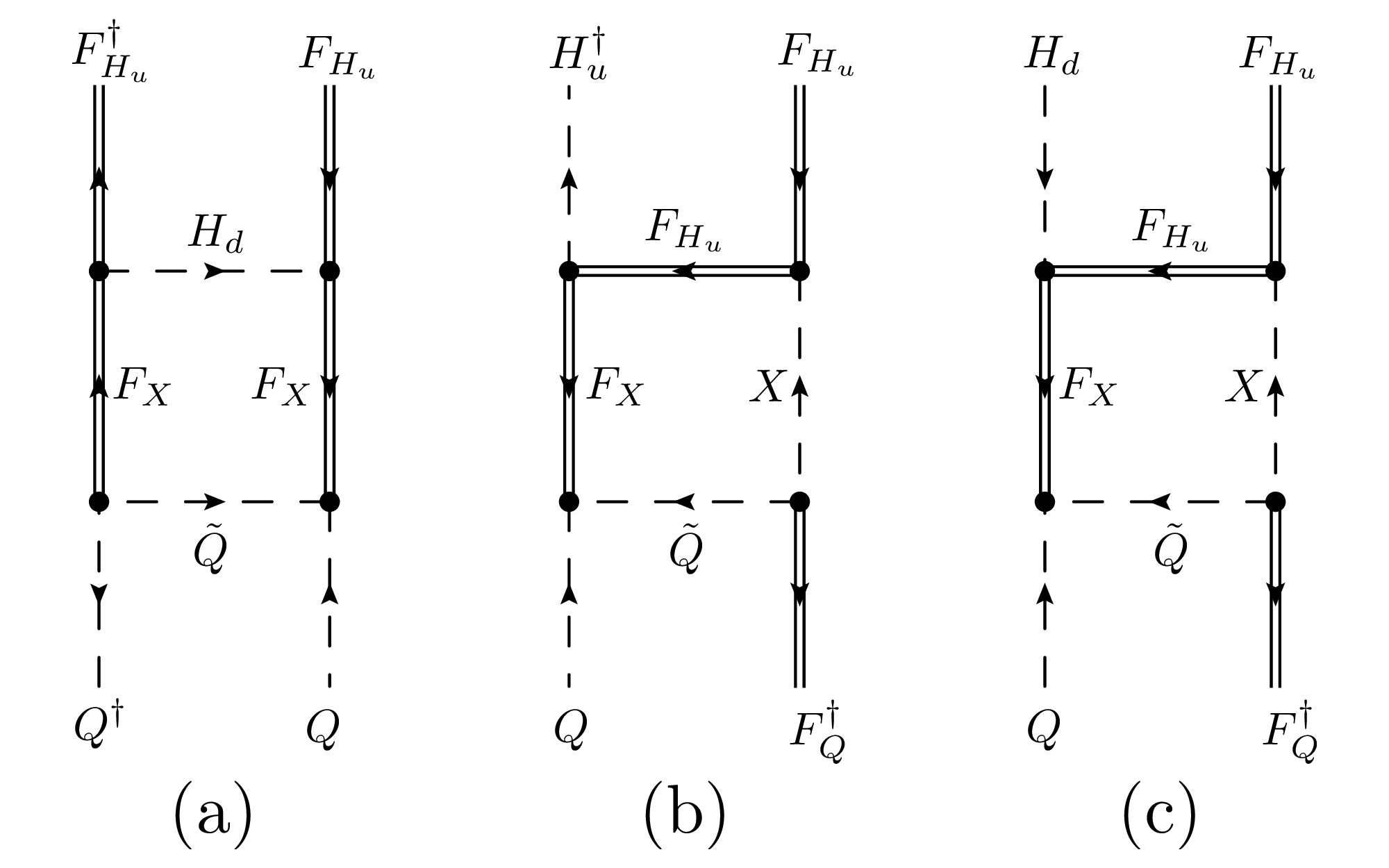}} 
 
  In perturbation theory, the renormalization of $c_{m_{u,d},L}$ proportional to $|c_\mu|^2$ or $|c_{A_{u,d}}|^2$ arises at one loop and $\CO(\epsilon / N_c)$. We may compute these loop diagrams in components using a suitably clever choice of external lines. Focusing on $c_{m_{u,d},L} Q^\dag Q F_{H_{u,d}}^\dag F_{H_{u,d}} $, for example, there is one diagram proportional to $|c_\mu|^2$, corresponding to the first diagram shown in \bzdiagrams .  Similarly, for $c_{m_{u,d},L} F_Q^\dag Q H_{u,d}^\dag F_{H_{u,d}} $ there is one diagram proportional to $|c_{A_{u,d}}|^2$, corresponding to the second diagram in \bzdiagrams . In contrast, the renormalization of $c_{m_{u,d},J^X}$ first arises at two loops and $\CO(\epsilon^2 / N_c^2)$. Similarly, the renormalization of $c_{B_\mu,L}$ proportional to $c_\mu (c_{A_u}^\ast + c_{A_d}^\ast)$ arises at one loop and $\CO(\epsilon / N_c)$. For the external components $c_{B_\mu, L} F_Q^\dag Q F_{H_u} H_d$ there is one diagram proportional to $c_\mu c_{A_u}^\ast$, corresponding to the third diagram in \bzdiagrams , while for $c_{B_\mu, L} F_Q^\dag Q F_{H_d} H_u$ there is an analogous diagram proportional to $c_\mu c_{A_d}^\ast$. As was the case for $c_{m_{u,d},J^X}$, the renormalization of  $c_{B_\mu, J^X}$ first arises at two loops and $\CO(\epsilon^2 / N_c^2)$, and we do not consider it here.
  
Computing these one-loop diagrams in $\overline{MS}$, the counterterms cancelling UV divergences associated with the one-loop diagrams renormalizing $c_{m_{u,d},L}$ and $c_{B_\mu,L}$ yield contributions to the beta functions of the form
\eqn\cmLbeta{\eqalign{
  {d c_{m_{u,d},L}\over d\log\Lambda}  &\supset -2 \sqrt{2 \over 3} {\epsilon \over N_c} (|c_\mu|^2+|c_{A_{u,d}}|^2) + \dots \cr
 {d c_{B_\mu,L} \over d\log\Lambda}  &\supset -2 \sqrt{2 \over 3} {\epsilon \over N_c} (c_\mu c^\ast_{A_u} + c_\mu c^\ast_{A_d}) + \dots 
}}
in agreement with \betagenii. This directly confirms the hidden sector renormalization calculated using superconformal perturbation theory in \betagen\ via standard perturbation theory to one loop at the free fixed point.

\subsec{Confirming the field redefinition argument}

It is also straightforward to see that this toy model is also consistent with the results expected from field redefinitions in the UV. The validity of the field redefinition argument requires the beta functions to take the form
\eqn\betagensimple{\eqalign{
 {d\over d\log\Lambda} \pmatrix{ c_{m_{u,d},L}\cr c_{m_{u,d},J^X}} 
&\approx \Gamma \pmatrix{ c_{m_{u,d},L}\cr c_{m_{u,d},J^X}}
 -(\Gamma - 2\D_X \times{\bf 1}) \pmatrix{ 0\cr 1} (|c_\mu|^2 +|c_{A_{u,d}}|^2) \cr
  {d\over d\log\Lambda} \pmatrix{ c_{B_\mu,L}\cr c_{B_\mu,J^X}} 
&\approx \Gamma\pmatrix{ c_{B_\mu,L}\cr c_{B_\mu,J^X}}
 - (\Gamma - 2\D_X\times{\bf 1})\pmatrix{ 0\cr 1} (c_\mu c^\ast_{A_u} + c_\mu c^\ast_{A_d})
}}
This field redefinition prediction agrees with the result from superconformal perturbation theory provided
\eqn\condition{
S^{-1} \pmatrix{ \CC_{\Delta_1} \cr \CC_{\Delta_2}} = \pmatrix{0 \cr 1}
}
We can check this directly in our toy model since \condition\ is precisely what is computed by the non-log-enhanced terms in the $X^\dagger-X-L$ and $X^\dagger-X-J^X$ three point functions. These terms are scheme-dependent starting at $\CO(\epsilon)$, so the only scheme-independent contributions come from tree-level diagrams in \opediagrams; these yield $0$ for $c_{m_{u,d},L}, c_{B_\mu,L}$ and $1$ for $c_{m_{u,d}, J^X}, c_{B_\mu, J^X}$. Thus \condition\ is trivially satisfied, rendering explicit agreement between the expectations from superconformal perturbation theory, direct perturbative calculation, and field redefinitions in the UV.

\footatend\vfill\supereject\immediate\closeout\rfile\writestoppt
\baselineskip=14pt\centerline{{\bf References}}\bigskip{\frenchspacing%
\parindent=20pt\escapechar=` \input refs.tmp\vfill\eject}\nonfrenchspacing

\end